\documentclass{aastex}
\usepackage{natbib}
\usepackage{emulateapj5}

\def\kms{\,{\rm km}\,{\rm s}^{-1}}

\newcommand{\FP}{Fokker-Planck}

\newcommand{\rc}{\mbox {$r_{\rm c}$}}

\newcommand{\trh}{\mbox {$t_{\rm rh}(0)$}}
\newcommand{\tcc}{\mbox {$t_{\rm cc}$}}
\newcommand{\tdis}{\mbox {$t_{\rm dis}$}}

\newcommand{\fb}{\mbox {$f_{\rm b}$}}
\newcommand{\go}{\gtrsim}
\newcommand{\lo}{\lesssim}

\newcommand{\gaoetal}{GGCM91}



\begin{document}

\shorttitle{MONTE CARLO SIMULATIONS OF GLOBULAR CLUSTER EVOLUTION. III.}
\shortauthors{FREGEAU, G\"URKAN, JOSHI, \& RASIO}
\submitted{Accepted for publication in ApJ}

\title{Monte Carlo Simulations of Globular Cluster Evolution. III. Primordial Binary Interactions}

\author{J.\ M.\ Fregeau\altaffilmark{1}, M.\ A.\ G\"urkan\altaffilmark{2}, 
	K.\ J.\ Joshi\altaffilmark{3}, \& F.\ A.\ Rasio\altaffilmark{4}}

\altaffiltext{1}{Department of Physics, MIT 37-624A, 77 Massachusetts Ave, 
                 Cambridge, MA 02139; {\tt fregeau@mit.edu}}
\altaffiltext{2}{Department of Physics and Astronomy, Northwestern University; {\tt ato@northwestern.edu}}
\altaffiltext{3}{IBM Corporation, 404 Wyman Street, Waltham, MA 02454; {\tt kjoshi@alum.mit.edu}}
\altaffiltext{4}{Department of Physics and Astronomy, Northwestern University; {\tt rasio@northwestern.edu}}

\begin{abstract}
We study the dynamical evolution of globular clusters using our 2D Monte Carlo code 
with the inclusion of primordial binary interactions for equal-mass stars.
We use approximate analytical cross sections for energy generation from binary--binary 
and binary--single interactions. After a brief period of slight contraction {\em or expansion\/}
of the core over the first few relaxation times, all clusters enter a much longer phase of 
stable ``binary burning'' lasting many tens of
relaxation times. The structural parameters of our models during this phase match well those 
of most observed globular clusters. At the end of this phase, clusters that have survived tidal
disruption undergo deep core collapse, followed by gravothermal oscillations.
Our results clearly show that the presence of even a small fraction of binaries in a cluster
is sufficient to support the core against collapse significantly beyond the normal core collapse
time predicted without the presence of binaries. For tidally truncated systems, collapse is easily 
delayed sufficiently that the cluster will undergo complete tidal disruption before core collapse. 
As a first step toward the eventual goal of computing all interactions exactly using 
dynamical three- and four-body integration, we have incorporated an 
exact treatment of binary--single
interactions in our code. We show that results using analytical cross sections are in 
good agreement with those using exact three-body integration, even for small binary 
fractions where binary--single interactions are energetically most important. 
\end{abstract}


\section{Introduction}

The realization about 10 years ago that primordial binaries are
present in globular clusters in dynamically significant numbers has completely
changed our theoretical perspective on these systems
\citep[see, e.g., the review by][]{1992PASP..104..981H}. Most importantly,
dynamical interactions between hard primordial binaries and other single
stars or binaries are thought to be the primary energy generation mechanism 
responsible for supporting a globular cluster against core collapse 
\citep{1989Natur.339...40G, 1990ApJ...362..522M, 1991ApJ...372..111M, 1991ApJ...370..567G}. 
The term ``binary burning'' is now often used
by analogy with hydrogen burning for stars. In the same way that hydrogen burning allows a star like 
the Sun to remain in thermal equilibrium on the main sequence for a time much longer than the Kelvin-Helmholtz
timescale, primordial binary burning allows a globular cluster to maintain itself in quasi-thermal 
equilibrium and avoid core collapse for a time much longer than the two-body relaxation timescale.

In addition, strong dynamical interactions
involving binaries can explain very naturally the large numbers of exotic objects found in 
dense star clusters. Exchange interactions between hard primordial binaries and 
neutron stars inevitably produce
large numbers of X-ray binaries and recycled pulsars in
globular clusters 
\citep{1991A&A...241..137H, 1995ApJS...99..609S, 1998MNRAS.301...15D, 2000ApJ...532L..47R}.
Resonant interactions of primordial binaries
result in dramatically increased collision rates for main-sequence stars
in globular clusters and even open clusters 
\citep{1996MNRAS.281..830B, cheung2003, 1989AJ.....98..217L, 1992AJ....103.1928L}.
Direct observational evidence for stellar 
collisions and mergers of main-sequence stars in globular
clusters comes from the detection of large numbers of bright blue stragglers
concentrated in the dense cluster cores
\citep{1995ARA&A..33..133B, 2002AJ....123.2541B, 2001ApJ...561..337F}.
Previously it was thought that primordial binaries
were essentially nonexistent in globular clusters, and so other mechanisms
such as tidal capture and three-body
encounters had to be invoked in order to form binaries dynamically during
deep core collapse. However, these other mechanisms have some serious
problems, and are much more likely to result in mergers than in the
formation of long-lived binaries 
\citep{1996IAUS..174..263C, 1992ApJ...385..604K, 1996ApJ...466..946K, 2002ApJ...576..899P, 1999A&A...347..123K, 1998ApJ...495..786K, 1993ApJ...409..617L}.
Multiple mergers of main-sequence 
stars and runaway collisions in young star clusters could lead to the formation of 
a massive central black hole 
in some systems 
\citep{1993ApJ...418..147L, 2002ApJ...578L..41G, 2002ApJ...576..899P}.

The primordial binary fraction is therefore a key input parameter for any
realistic study of dense star cluster dynamics \citep{1992PASP..104..981H}. Early
determinations of binary fractions in globular clusters came from
observations of spectroscopic binaries with red giant primaries 
\citep[see, e.g., \citealt{1996AJ....112..574C} for a more recent study]{1988AJ.....96..123P}
as well as eclipsing binaries \citep{1990AJ....100..469M, 1994AJ....108.1810Y}.
Hubble Space Telescope observations have provided direct constraints
on primordial binary fractions in the central regions of many globular
clusters, where binaries are expected to concentrate because of mass
segregation. \citet{1997ApJ...474..701R} used observations of a broadened
main sequence in NGC 6752 to derive a binary fraction in the range
15\%--38\% for the inner cluster core. Their method has now been applied to
many other clusters. For example, \citet{2002AJ....123.2541B} derive a similar
binary fraction, in the range 0.08--0.38, in the central region of NGC 288.
Adding proper motion information can lead to much tighter constraints, as in
the case of NGC 6397, where \citet{2002scmc.conf..163C} derive a binary fraction
$\lo 5-7$\% near the center.

Despite the crucial role of primordial binaries in the dynamical
evolution of a dense star cluster, the overall
evolution of the binary population within a cluster, and its direct implications
for the formation rate of observable systems such as recycled pulsars and blue stragglers, 
remains poorly understood theoretically.
One reason is that the relative importance of binary interactions in a cluster, like many
other dynamical processes, depends in a complex manner on the number of stars in the system.
This makes it difficult to extend results obtained from small direct $N$-body
simulations to realistic globular cluster models. In particular, the rate at which 
binaries are ``burned'' and, ultimately, destroyed or ejected from the cluster depends
on the size of the cluster.
When the initial primordial binary fraction is below a certain critical value,
a globular cluster core can run out of binaries before the end of its lifetime,
i.e., before disruption in the tidal field of the Galaxy \citep{1994ApJ...427..793M}.
Without the support of binaries, the cluster will then undergo a
much deeper core collapse, perhaps followed by gravothermal oscillations
\citep{1983MNRAS.204P..19S, 1994ApJ...421..195B, 1996ApJ...471..796M}.
At maximum contraction, the core density may increase by many orders 
of magnitude, leading to greatly enhanced interaction rates.

Detailed numerical studies of globular cluster evolution with primordial binaries are
still lacking, for several reasons. First, the inclusion of even
a modest fraction of primordial binaries adds a very significant
computational overhead to $N$-body simulations. This is mainly due
to the extra computations required to treat binary interactions, 
but also because the lifetime of a cluster 
can be significantly extended (by up to many orders of magnitude) 
through binary burning. In addition, in direct $N$-body simulations,
the extremely large ratio of the overall cluster dynamical time to the orbital 
period of close binaries (as large as $\sim 10^{10}$ in a globular cluster!)
introduces many computational difficulties. This makes $N$-body 
simulations with primordial binaries prohibitively expensive for
$N \gtrsim 10^4$ stars, although special-purpose supercomputers such as the
new GRAPE-6 may increase this limit in the near future \citep{2001dscm.conf...87M}.
Orbit-averaged calculations, like direct 
\FP\ integrations and Monte Carlo simulations, get around 
this problem by treating binaries just like 
single stars, except during brief periods of strong interactions.
Unfortunately, this requires that cross sections for strong 
interactions involving binaries be known accurately, for a 
wide range of binary parameters (masses, semi-major axes, and 
eccentricities). These cross sections are difficult 
to determine in general, and reliable semi-analytic fits to 
numerical scattering experiments are only available for 
simple configurations such as those involving equal-mass stars.
For these reasons, most previous numerical studies of globular clusters with primordial binaries
have been limited either to clusters with equal-mass stars
\citep{1991ApJ...370..567G, 1980ApJ...241..618S}, or to very small clusters with
$N \sim 10^3 - 10^4$ stars 
\citep{1992MNRAS.257..513H, 2001MNRAS.323..630H, 1990ApJ...362..522M, 1991ApJ...372..111M, 1994ApJ...427..793M}.
Simplified treatments have also been employed in which the
dynamics of the binaries was followed in a static cluster background 
\citep{1992ApJ...389..527H} or in a background cluster modeled as an evolving gas sphere
\citep{2000MNRAS.317..581G}.

The results of \citet[hereafter GGCM91]{1991ApJ...370..567G}, 
based on direct \FP\ integrations,
were the first to clearly illustrate 
the dominant effect of even a small fraction of primordial binaries 
on the evolution of a globular cluster. In this paper, we present the first 
study of globular cluster evolution with primordial binaries based on
self-consistent Monte Carlo simulations with a realistically large number of stars ($N\go 10^5$).
Partly in order to allow better 
comparison of our results with those of \gaoetal, we use similar initial 
conditions and cross sections for binary--binary and binary--single 
interactions, even though our method for implementing these cross 
sections in the Monte Carlo scheme is completely different. 
In addition, the results of \gaoetal\ were obtained using
a 1-D \FP\ method (in which isotropy in velocity space is enforced). 
More realistic 2-D (anisotropic) \FP\ calculations with 
primordial binaries have never been reported in the literature, 
to the best of our knowledge. Even for the 1-D 
calculations, and with only a single 
parameter representing the internal structure of binaries 
(namely, their binding energy), the inclusion of binary--binary
interactions significantly increased the overall computation time. 
Since the \FP\ method uses distribution functions to represent the 
system, every new parameter adds a new dimension to the phase space, 
making the \FP\ equation more difficult to solve numerically. 
It has also been shown recently that the 
1-D treatment is inadequate in dealing with some aspects of 
the evolution, such as the escape rate from tidally truncated
clusters 
\citep{1998ApJ...503L..49T, 2000ApJ...535..759T}.
Many difficulties in the direct \FP\ approach come from the basic representation
of the system in terms of smooth distribution functions. Neglecting the
discrete nature of the system makes it impossible to follow 
the details of individual interactions, such as binary--single or
binary--binary interactions. The implicit
assumption that $N \to \infty$ also makes it difficult to scale the results 
of direct \FP\ simulations to finite systems
with different numbers of stars.

Our Monte Carlo method provides an intermediate approach, which
combines many of the benefits of direct $N$-body simulations (such as the description 
of the cluster on a star-by-star basis and the possibility to treat each individual 
interaction in detail) with the
speed of an orbit-averaged calculation.
Our method is also 2-D in velocity space
by construction, and hence properly accounts for any velocity
anisotropy in the system. Another benefit of the method is that
it allows a wide range of binary parameters to be used without
having to modify the underlying orbit-averaged calculation of the
relaxation processes. In principle, individual interactions 
can be treated in as much detail as in direct $N$-body simulations,
by computing all strong encounters exactly using three-body or four-body
integrators. 
As a first step, for this paper, we have incorporated a three-body dynamical integrator
into our code, which allows binary--single interactions to be computed
exactly (without reference to approximate, pre-compiled cross sections). 
This allows us to follow the outcomes of interactions more precisely, 
and, most importantly, will allow us in the future to extend our code to multi-mass 
systems, for which analytic cross sections are not available.

\section{Treatment of Binary Interactions}\label{sec:binint}

We use the basic H\'enon-type Monte Carlo method for modeling the dynamical 
evolution of clusters
as a sequence of equilibrium models subject to regular velocity perturbations
\citep{1971Ap&SS..13..284H, 1971Ap&SS..14..151H}; our code has been described 
in detail by \citet[hereafter Papers I and II]{2000ApJ...540..969J, 2001ApJ...550..691J}.
The regular velocity perturbations are calculated using H\'enon's method to
represent the average effect of many long-range small-angle gravitational scattering
encounters using one suitably chosen encounter with a nearby star \citep{1971Ap&SS..14..151H}.
At each time step, we calculate the Monte-Carlo realized radial position and velocity
of each star (assuming spherical symmetry), which we use to calculate whether
two objects (binary-single or binary-binary) will interact strongly.  These strong
interactions are performed using either simple recipes based on cross sections,
or a dynamical integrator.  For most of the work reported here,
we use cross sections for the treatment of close binary--binary and binary--single interactions.
These cross sections were compiled from analytic fits to the results of numerical scattering 
experiments available in the literature. Given
the very large parameter space, reliable cross sections are available only for
equal-mass encounters, and so we study only single-component clusters in this paper.
All single stars are assumed to have the same mass, and all binaries contain two identical
stars with the same mass as the background single stars. All stars are treated as point masses,
i.e., we neglect physical collisions between stars during interactions 
\citep[cf.][]{1996MNRAS.281..830B, cheung2003}.
Our implementation follows closely that
used in the \FP\ study by \gaoetal, which will serve as the main comparison
for our work.

\subsection{Units and Definitions}\label{sec:units}

In our code we use the system of units defined by setting $G=M_0=-4E_0=1$, where
$M_0$ is the initial cluster mass, and $E_0$ is the initial cluster energy
(excluding the binding energy in binaries).  The corresponding unit of time is then
$t_{\rm dyn}(0) = G M_0^{5/2} (-4E_0)^{-3/2}$.  However, the natural timescale
for cluster evolution is the relaxation time
\begin{equation}
\label{eq:trlx}
  t_r(0) \equiv \frac{N_0}{\ln(\gamma N_0)} t_{\rm dyn}(0) \, ,
\end{equation}
where $\gamma$ is a constant of order unity that must be determined experimentally.
This relaxation time is used as the time unit in the Monte Carlo code. Therefore the
$\ln(\gamma N_0)$ dependence factors out from all expressions used in simulating
two-body relaxation 
\citep[see][or Paper I]{1971Ap&SS..14..151H}.
When reporting results, however, we scale all times to the
initial half-mass relaxation time, which we calculate using the standard definition given by 
\citet{1971ApJ...164..399S},
\begin{equation}
\label{eq:trh}
  t_{\rm rh}(0) \equiv \frac{0.060 N_0}{\log_{10}(\gamma N_0)} \left(\frac{r_{\rm h}^3(0)}{GM_0}\right)^{1/2}
  \equiv \frac{0.138 N_0}{\ln(\gamma N_0)} \left(\frac{r_{\rm h}^3(0)}{GM_0}\right)^{1/2} \, ,
\end{equation}
where $r_{\rm h}(0)$ is the initial half-mass radius of the cluster. 
Since $t_{\rm dyn}(0) \simeq [r_{\rm h}^3(0)/(GM_0)]^{1/2}$,
we have $t_{\rm rh}(0) \sim 0.1 t_r(0)$, where the numerical coefficient depends
on the value of $r_{\rm h}(0)$ for the particular initial model.

When calculating rates for processes that do not occur on the relaxation timescale,
such as dynamical interactions, one 
must adopt a specific value for $\gamma$ in the Coulomb logarithm
when converting between dynamical and relaxation
times.  In all our simulations for this paper we use $\gamma=0.4$, the standard value adopted in most
previous \FP\ simulations, including those of \gaoetal\ (cf.\ Paper~I, where we show that $\gamma\simeq 0.1$
provides the best agreement with direct $N$-body simulations for the evolution of a single-component
cluster to core collapse).  In addition,
in calculating the binary--binary and binary--single interaction rates, 
\gaoetal\ use $N_{\rm c}$, the {\em current\/}
number of stars in the cluster core, instead of $N_0$ in the denominator of eq.~\ref{eq:trlx} 
when converting between dynamical and relaxation times. Although there is no rigorous justification
for this choice, it appears reasonable, since the interactions occur mainly inside the high-density
core, and we adopt the same prescription in our simulations.

To estimate core quantities, including the number of stars in the core, we first
sample over a small number of stars, typically 0.1--1\% of the total number
of stars in the cluster, to calculate the central density, $\rho_0$, 
and velocity dispersion, $\sigma_0$.  Since the velocity dispersion varies
so slowly away from the center, we estimate the core velocity
dispersion as $\sigma_{\rm c} \simeq \sigma_0$.  The core radius
is then defined to be
\begin{equation}
  \label{eq:rc}
  r_{\rm c} = \left(\frac{3\sigma_{\rm c}^2}{4\pi G\rho_0}\right)^{1/2} \, ,
\end{equation}
and the number of stars in the core is calculated as
\begin{equation}
  N_{\rm c} = \frac{4\pi r_{\rm c}^3 \rho_0}{3 \bar{m}} = \frac{2\pi r_{\rm c}^3 \rho_{\rm c}}{3 \bar{m}} \, ,
\end{equation}
where $\bar{m}$ is the average mass of a star in the cluster, and
$\rho_{\rm c} \simeq 0.5 \rho_0$ 
\citep{1987degc.book.....S}.

\subsection{Binary--Single Interactions}\label{sec:binsingle}

In a single time step, the probability that a binary will strongly 
interact with another object (single or binary) is given by
\begin{equation}
\label{eq:nsigmav}
  P = \sigma w n \Delta t \, ,
\end{equation}
where $\sigma$ is the cross section for the interaction,
$w$ is the relative velocity at infinity, $n$ is the local number density of
stars (single or binary), and $\Delta t$ is the time step.  For binary--single
interactions, $\sigma=\sigma_{\rm bs}$, the binary--single interaction cross section, 
and $n=n_s$, the local number density of single stars.
In our code we calculate $n_s$ using a local sampling procedure,
and take $w$ to be the relative velocity between the nearest
single star and the binary.
The total cross section for close binary--single interactions
is computed as $\sigma_{\rm bs} = \pi b_{\rm max}^2$. Here $b_{\rm max}$ is the 
impact parameter which gives a distance of closest approach
between the binary and the single star of $r_{\rm min} = 3.5\, a$, where
$a$ is the binary semi-major axis. For a binary of mass $m_b$ and
single star of mass $m$ we have
\begin{equation}
\label{eq:bmax}
  b_{\rm max}^2 = r_{\rm min}^2\left(1 + \frac{2G(m+m_b)}{w^2 r_{\rm min}}\right) \, .
\end{equation}
The coefficient of 3.5 is chosen such that all interactions with a distance of closest 
approach greater than $r_{\rm min}$ result in only negligible 
energy transfer from the binary to the passing star in 
a fly by \citep[see, e.g.,][]{1975MNRAS.173..729H}.
As long as it is sufficiently large, the precise value of the coefficient has
very little influence on the results.

The binary--single interaction is performed only if a uniform 
random number between~0 and~1 is less than the computed probability. 
The interaction can in principle be 
computed exactly using a three-body dynamical integrator; this approach 
has many benefits, especially in providing 
an accurate way of distinguishing between the various possible
outcomes (see below). However, it also requires significantly more
computational resources than using a simple analytic prescription. 
Following \gaoetal, in these equal-mass simulations, we assume that the only outcome
is binary hardening, and we use a
semi-analytic fit 
\citep[eq.~{[}6-27{]}]{1987degc.book.....S}
to numerical results
\citep{1984ApJS...55..301H}
to compute the translational energy released. 
Let $y = \Delta \epsilon / \epsilon$
be the fraction of the binding energy of the binary that is 
released as translational energy. The differential cross 
section for the interaction is given by 
\citep{1991ApJ...370..567G},
\begin{equation}
\label{eq:dsigmabs}
\frac{d\sigma_{\rm bs}}{dy} = 12.48 \pi a^2 \left(\frac{w}{v_{\rm cr}}\right)^{-2}(1+y)^{-4} y^{-0.5}\, ,
\end{equation}
\noindent
where the critical velocity is $v_{\rm cr} = (3Gm/2a)^{1/2}$, the velocity
at infinity for which the total energy of the system is zero and 
complete ionization is possible.  The quantity $y$ is drawn randomly from
eq.~(\ref{eq:dsigmabs}) using the rejection technique.  With this
recoil energy and a scattering angle drawn at random in the center-of-mass
frame, new velocities and orbital energies
in the cluster are calculated for the emerging binary and single star.

As a first step toward the eventual goal of treating all binary interactions
exactly, we have incorporated into our code a dynamical integrator
to perform binary--single interactions. Specifically, we use the three-body 
integrator {\tt scatter3} from the {\tt Starlab} software
environment 
\citep[see Appendix~B of][and {\tt http://www.manybody.org}]{2001MNRAS.321..199P}.
{\tt Scatter3} uses a time-symmetrized Hermite integrator and analytical continuation of 
unperturbed orbits to evolve the three-body system until an unambiguous outcome
is obtained.  The main benefit of using an exact treatment is the ability to study 
non-equal-mass systems, 
although for comparison with cross sections we restrict ourselves to the
equal-mass case here.  The implementation of the three-body integrations
follows that of cross sections: first, the probability for an encounter to occur is calculated
according to eqs.~(\ref{eq:nsigmav}) and (\ref{eq:bmax}); next, with velocity
at infinity $w$, the impact parameter $b$ is
chosen randomly in area, i.e., with probability $dP(b) = 2\pi b\, db / (\pi b_{\rm max}^2)$.
The binary eccentricity is assumed to follow a thermal distribution with $dP(e)=2e$, and all
angles are chosen assuming random orientation and phase.
The dynamical interaction is then calculated
and its outcome is used to determine the new binding energy of the binary and the new orbits
for the binary and single star in the cluster.
The current implementation properly handles the outcomes of preservation
and exchange, but, for the sake of comparison with cross sections, currently ignores ionizations.
This is justified here since we consider only hard primordial binaries in our simulations
(see Sec.~\ref{sec:initcond}).

\subsection{Binary--Binary Interactions}

To calculate the probability that a close binary--binary interaction should
occur in a time step, we use eq.~(\ref{eq:nsigmav}) with
$\sigma=\sigma_{\rm bb}$, the binary--binary interaction cross section, 
and $n=n_b$, the local number density of binaries.
We calculate $n_b$ using a local sampling procedure,
and we take $w$ to be the relative velocity between the current binary
and the nearest binary.  
Following \gaoetal, for the binary--binary interaction cross section we use the results of 
\citet{1983MNRAS.203.1107M, 1983MNRAS.205..733M, 1984MNRAS.207..115M, 1984MNRAS.208...75M}
for encounters between equal-mass binaries. In the case where one binary has a 
much higher binding energy than the other ($\epsilon_1 \gg \epsilon_2$), 
\citet{1984MNRAS.207..115M} provides a semi-analytic fit to his numerical 
results, giving a collision cross section
\begin{equation}
\label{eq:sigmabb}
\sigma_{\rm bb} \approx 16.6\left[\ln\left(\frac{29 |\epsilon_2|}{m w^2 + 0.04|\epsilon_2|}\right)\right]^{2/3}
		\frac{G m a_2}{w^2} \, ,
\end{equation}
\noindent
where $w$ is the relative velocity of the two binaries at infinity, $m$ is
the mass of each star in the binaries, and $a_2$ is the semi-major axis of
the softer binary. 

An interaction between two binaries can result in 
many possible outcomes.  Since we consider only hard binaries
in this study, the most probable
outcomes are (1) disruption of the softer binary and 
hardening of the harder one, and (2) the formation of a 
stable hierarchical triple with a single star ejected. As much as $\sim1/3$ of 
close binary--binary encounters
may result in the formation of such a triple.
However, for a triple system to remain long-lived
in the dense cluster environment, its outer orbit must be
sufficiently tight. The formation of a {\em long-lived\/} hierarchical triple
is expected to be much less common in a dense cluster 
\citep[see][]{2000ApJ...528..336F}.
Therefore we assume for simplicity that all hierarchical triples formed
are disrupted immediately. The only outcome of
binary--binary encounters that we treat in our simulations is then case (1) above.
\citeauthor{1984MNRAS.207..115M} finds that, on
average, approximately one half of the combined binding energy 
($\epsilon_1 + \epsilon_2$) of the two binaries is released in 
the form of translational energy, $\Delta E_t = y (\epsilon_1 + \epsilon_2)$. 
The semi-analytic fit given by \citet{1984MNRAS.207..115M} for the distribution
of translational energies produced is rather complicated. Instead,
we use a simplified version of the distribution, 
\begin{equation}\label{eq:G(y)}
G(y) = \frac{49}{4} y \left(1+\frac{7}{2}y^2\right)^{-11/4} \, ,
\end{equation}
proposed by \gaoetal. 
The mean value of this distribution, $\langle y\rangle \simeq 0.47$,
is in good agreement with the results of \citeauthor{1984MNRAS.207..115M} for interactions
resulting in a binary and two single stars. 

We also adopt a simplified overall binary--binary 
collision cross section, by replacing the expression in 
square brackets in eq.~(\ref{eq:sigmabb}) by its value at $\epsilon_2 = \frac{1}{2} m w^2$,
yielding
\begin{equation}
\sigma_{\rm bb} = 31.8 \frac{G m a_2}{w^2} \, .
\end{equation}
\noindent
The energy required to disrupt the softer binary, as well as the total 
translational energy released in the collision $\Delta E_t$, are both 
generated at the expense of the surviving binary. Thus
the binding energy of the surviving pair increases by
an amount $\epsilon_2 + y(\epsilon_1 + \epsilon_2)$. 
According to \citet{1983MNRAS.203.1107M}, for collisions between
binaries of equal binding energies producing a binary and two 
single stars, typically about $1/4$ of the translational energy produced is
carried away by the binary, and the remaining is distributed 
randomly among the two single stars. For simplicity,
we assume that this prescription is applicable to collisions 
between binaries of unequal binding energies as well.
We select the direction of the recoil velocity between the binary
and the single stars randomly in the center-of-mass frame.

If a binary does not undergo a strong interaction with a single
star or another binary, it is then treated as a single star in 
the usual two-body relaxation step (see Paper~I), during which its internal
structure is left unchanged.

\section{Results}

\subsection{Initial Conditions and Summary of Model Results}\label{sec:initcond}

For our initial cluster models we use both the Plummer model, assumed to be isolated (i.e.,
with no tidal boundary enforced), and a variety of tidally truncated King models, assumed 
to be on a circular
orbit in the Galaxy (i.e., with a fixed external tidal potential). Mass loss through the tidal 
boundary is treated as in Paper~II, using a criterion based on the apocenter distance of each 
stellar orbit in the cluster, and an iterative procedure to determine both the mass loss 
and the new position of the tidal boundary after each relaxation timestep. 
The initial binary fraction $\fb$ (defined as the 
fraction of stars, by number, that are binaries) varies between 0 and 30\%. In a few cases, for
calibration, we have also performed simulations in which the binaries are present, but all
interactions are turned off; these models are equivalent to two-component models in which a
small fraction of (single) stars have twice the mass of the background stars 
(see \citealt{2000ApJ...539..331W} and \citealt{2002ApJ...570..171F} for other studies 
of two-component clusters using our code).

The binaries are distributed initially in the cluster 
according to the same density profile as for single stars. Hence no initial
mass segregation is assumed for the binaries.  The distribution
of the internal binding energy of the binaries is assumed to be uniform in
$\log \epsilon$ between a minimum value $\epsilon_{\rm min}$ and a maximum value
$\epsilon_{\rm max}$. Following \gaoetal, we consider only hard binaries,
with the minimum binding energy $\epsilon_{\rm min} = m \sigma_{\rm c}(0)^2$, where $\sigma_{\rm c}(0)$ 
is the initial
central velocity dispersion. Soft binaries, if present, would be assumed to be ionized
(destroyed) as soon as they participate in a strong interaction. Therefore they would not 
affect the overall evolution of the cluster significantly. For the maximum binding 
energy we take $\epsilon_{\rm max} = 133 \epsilon_{\rm min}$, which is approximately
the binding energy of a contact binary for two solar-like stars if $\sigma_{\rm c}(0) \simeq 10\, \kms$.
The precise value of $\epsilon_{\rm max}$ has little influence on our results, since
very hard binaries behave essentially as single more massive stars (with very small 
interaction cross section). 

Table \ref{tab:1} lists the parameters of the main models we considered, as well as the main results 
of our simulations for each cluster. 
The first column identifies the initial cluster model, Plummer or King, and the value of the
concentration parameter $W_0$ (dimensionless central potential) for King models. The second column
gives the initial binary fraction $\fb$. All simulations were performed with $N=3\times10^5$ stars
(including binaries) initially in the cluster.
The following columns summarize the main results of our dynamical simulations.
For each model we first give the time of core collapse $\tcc$, in units of the 
initial half-mass relaxation time $\trh$, defined by eq.~(\ref{eq:trh}). Here core collapse
is defined as the moment when the core density reaches its {\em first maximum\/}. 
This can be determined typically to within a statistical error of at most a few percent
in our simulations. We then give 
the total mass of the cluster at the moment of core collapse (in units of its initial
total mass), and the fraction of binaries that remain at that moment. For clusters that
disrupt completely before reaching core collapse, we list the disruption time $\tdis$
instead of $\tcc$. 

\subsection{Comparison with Direct Three-Body Integration}\label{sec:comp}

As a simple test of our code and the approximate treatment of interactions,
we compare the use of cross sections with dynamical integrations of
binary--single encounters. In future work, we will also implement dynamical integrations
of binary--binary interactions, and we will use more detailed comparisons to re-calibrate
the various recipes based on cross sections. Here our intent is merely to demonstrate that
these simple recipes are reasonably accurate.
We have not changed our prescriptions to try to better match the results of the dynamical integrations,
since a main goal in this first study is to provide comparisons with the \FP\ simulations
of \gaoetal\ that used the same simple prescriptions.
For this test {\em binary--binary interactions were turned off\/}. In reality,
they tend to dominate the energy production (see Sec.~\ref{sec:isoclus}). Thus this simple test also 
allows us to study specifically the effects of three-body interactions on the overall cluster 
evolution.

\begin{figure}[t]
\epsscale{0.74}
\plotone{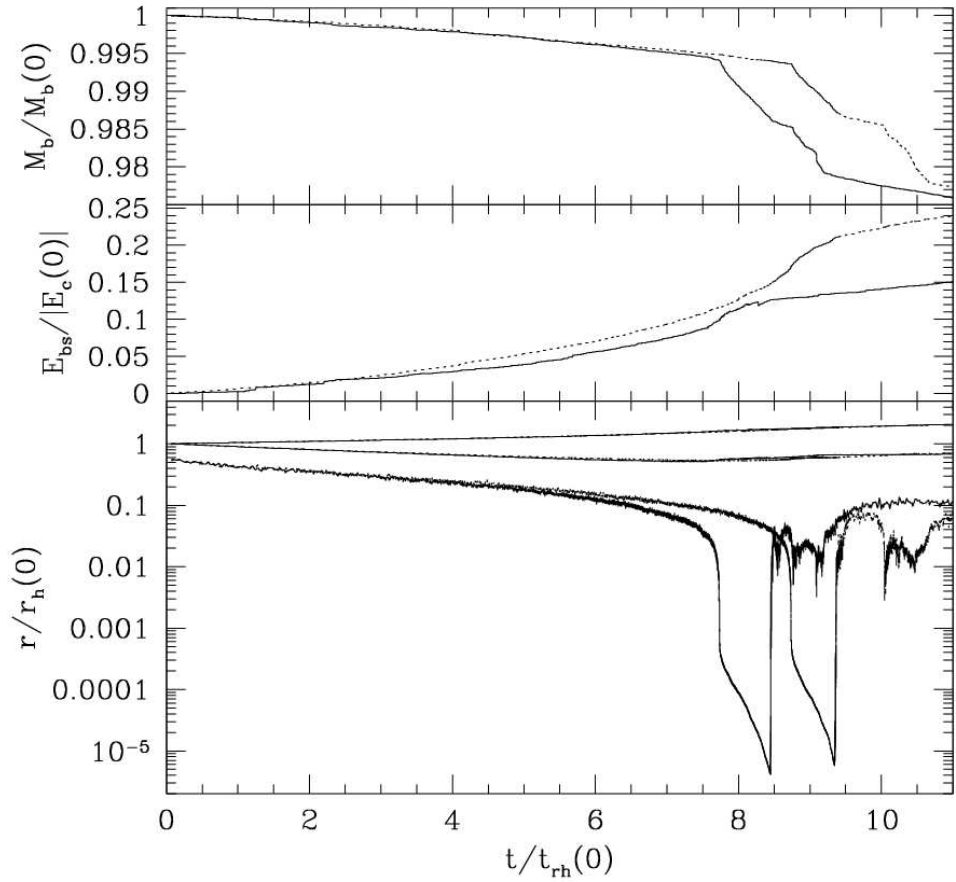}
\begin{center}
\caption{Comparison between the use of direct three-body integrations (solid lines) and cross
sections (dashed lines) in calculating the evolution of a Plummer model containing 
$N=3\times 10^5$ stars with 20\% binaries initially. In both cases  
{\em binary--binary interactions were turned off\/}.  The top panel
shows the total mass in binaries. The middle panel shows
the energy generated in binary--single interactions,
as a fraction of the total initial binding energy of the cluster.
The bottom panel
shows (from top to bottom) the half-mass radius of single stars,
the half-mass radius of binaries, and the core radius, in units of
the initial half-mass radius. Time is given in units of the initial 
half-mass relaxation time.  The agreement 
between the two methods is strong, although the energy production 
is slightly overestimated in the treatment based on cross sections, leading to divergent evolutions 
near core collapse (the model calculated with direct three-body integrations collapses earlier).
\label{fig:comp}}
\end{center}
\end{figure}

Figure~\ref{fig:comp} shows the evolution of
an isolated cluster described initially by a Plummer model with $N=3\times 10^5$ stars 
and 20\% binaries. 
Solid lines correspond to the simulation using direct three-body integrations, while dashed lines
show the results using our simple cross sections.
The top panel shows the total mass in binaries in the cluster, decreasing as binary burning proceeds.
Since all binaries in the model are hard, binary--single interactions (unlike binary--binary interactions)
cannot destroy a binary, and 
therefore binaries can only be lost by ejection from from the cluster (typically following significant
hardening through multiple interactions; see \citet{1992ApJ...389..527H}
and Sec.~\ref{sec:binev} below). The rate of
binary ejection accelerates abruptly at $t/\trh \simeq 8-10$ near core collapse.
The middle panel of Figure~\ref{fig:comp}
shows the energy generated in binary--single interactions,
as a fraction of the total initial binding energy of the cluster. By the time of core collapse,
this is only $\sim0.1$. This amount of energy is not sufficient to delay core collapse
significantly. In fact the binaries, through mass segregation, {\em accelerate\/} core collapse
in this (artificial) simulation (recall that the core collapse time of a single-component Plummer
model without binaries is given by $\tcc/\trh\simeq 14$).
The bottom panel shows various characteristic radii in the cluster:
from top to bottom, the half-mass radius of single stars,
the half-mass radius of binaries, and the core radius, all in units of
the initial half-mass radius.  

The agreement 
between the two methods is strong, although the total energy generated in
binary--single interactions is slightly smaller when calculated by direct dynamical integrations.
Consequently, 
the model using dynamical integrations reaches core collapse sooner 
than the model using cross sections, because less energy is generated
to support the core against collapse. 
We believe that this difference comes from the deterministic treatment
of binary hardening with cross sections, in which
every binary--single interaction 
results in a hardened binary. In reality the widest hard binaries in the simulation, which 
are right around the hard/soft boundary
(and have the largest interaction cross section) have roughly equal probabilities
of hardening and softening in an interaction \citep{1975MNRAS.173..729H}. 
To partly restore consistency between the two treatments, we
ignore dynamical integration outcomes that result in ionization of the
binary.  Were these included, the total energy generated in binary--single interactions
would decrease further, by roughly 50\%.  This would cause the binary population
to become more centrally concentrated.  Thus, in a realistic cluster simulation, 
we would expect the ratio of the energy generated by binary-single interactions 
to binary-binary interactions to decrease by more than 50\% compared to predictions
of cross-section based recipes.

\subsection{Isolated Clusters}\label{sec:isoclus}
We consider first the evolution of 
Plummer models containing $N=3\times 10^5$ stars with a range of binary fractions $f_{\rm b}$.
As a further test of our method, we show in Fig.~2 the evolution of the various 
energies and the virial ratio of 
a system with $f_{\rm b}=0.1$. Since dynamical relaxation is not built into our numerical method, 
the degree to which virial equilibrium is maintained during a simulation is our 
most important indicator of numerical accuracy. We monitor this, as well as energy conservation,
in all our runs, and terminate a calculation whenever these quantities deviate from their
expected values by more than a few percent (this typically happens when the number of binaries
has been reduced to a very small value, or, in tidally truncated clusters, when the total
number of stars remaining in the cluster becomes very small; See Sec.~\ref{sec:truncclus} below).

Figures 3, 4 and~5 show the evolution of models with $\fb=0.02$, 0.1, and 0.2, respectively.
The main impact of introducing binaries in the models is very clear: core collapse is delayed
considerably. Even for a cluster with only 2\% binaries initially (Fig.~3),
$\tcc$ increases by more than a factor 2. Clusters with $\fb \simeq 0.1 - 0.2$ 
can avoid core collapse for $\sim 100\,\trh$ (Figs.~4 and~5). For the vast majority
of globular clusters in our Galaxy, where $\trh \sim 10^9\,$yr, this timescale exceeds
a Hubble time. If all globular clusters in our Galaxy were born with $\fb \go 0.1$,
only those with very short initial relaxation times would have had a chance to reach core 
collapse. However, for real clusters, tidal truncation and mass loss (Sec.~\ref{sec:truncclus}) as well
as stellar evolution (Paper~II) complicate this
picture considerably.
 
In Figure~3, we also show for comparison the evolution of the core radius
for a model in which binaries are present but {\em all interactions are turned off\/}
(short-dashed line in the bottom panel). Even with a binary fraction as small as
$\fb =0.02$ in this case, core collapse occurs significantly earlier than 
in a single component Plummer model (at $t_{\rm cc}/\trh \simeq 10$ instead of 14).
This shows the expected tendency for the heavier component of binaries to {\em accelerate\/}
the evolution to core collapse, and the result is in good 
agreement with previous studies of core collapse
in two-component clusters \citep[see, e.g.,][]{2000ApJ...539..331W}.
Note that for sufficiently large binary fractions, these two-component models become
``Spitzer-unstable,'' i.e., the core collapse is driven entirely by the heavier component.
Using the stability criterion derived by \citeauthor{2000ApJ...539..331W}, 
$\Lambda \equiv (M_2/M_1) (m_2/m_1)^{2.4} < 0.32$,
here with an individual mass ratio $m_2/m_1 = 2$ and a total component mass ratio
$M_2/M_1 = 2f_{\rm b}/(1-f_{\rm b})$ we expect the Spitzer instability to appear whenever
the binary fraction $f_{\rm b} \go 0.03$. Thus all our models with binary fractions above a 
few percent should evolve on a relaxation timescale to a state where the dynamics of the
cluster core is largely dominated by the binaries. Indeed, looking at the middle panels
of Figures~4 and~5, we see 
that, with $f_{\rm b}=0.1-0.2$, the energy generation is largely dominated by binary--binary
interactions. In contrast, for $f_{\rm b}=0.02$ (middle panel of Fig.~3), binary--binary and
binary--single interactions contribute roughly equally.

To quantify the effect of primordial binary burning on the core collapse time, we have repeated
calculations with binaries present but all interactions turned off for six different models
with varying $\fb$. For each model, we can then properly calculate the ratio of core 
collapse times with and without binary interactions (but with mass segregation effects present in
both cases). 
The results are shown in Figure~6, where this ratio is plotted as a function of the binary
fraction. A simple linear fit gives
$$ \tcc \simeq \tcc(\fb=0) \times (75\,\fb + 1),$$ 
and reproduces the numerical results to within $\sim30$\% 
in the range $\fb = 0 - 0.3$. The notation we use here, ``$\tcc(\fb=0)$,''
means the core collapse time of a cluster with the same fraction of ``inactive'' binaries,
rather than with no binaries.

Also shown in Figure~3 for comparison is the result of a simulation in which all
interactions are included, but binary--single interactions are calculated by direct three-body
integrations as in Sec.~\ref{sec:comp} (long-dashed line in the bottom panel). This comparison is useful 
again as a test of the simple treatment based on cross sections, since binary--single
interactions play an important role as a source of energy in this model with $\fb=0.02$.
Our conclusion is the same as in Sec.~\ref{sec:comp}: the agreement is very good until $t/\trh\simeq 15$, 
but then the two simulations diverge and the model computed with direct three-body integrations
collapses slightly earlier ($\tcc/\trh \simeq 19$ instead of 22). In spite of this slight
offset, after core collapse
the core re-expansion and gravothermal oscillations also look very similar in the two 
simulations.

The top panels in Figures~3--5 show the evolution of the total cluster mass,
as a fraction of the initial mass, and the remaining mass in binaries, as a fraction
of the initial mass in binaries (this is also the remaining fraction by number
since all binaries have the same mass). Binary--binary interactions are the main process
responsible for the destruction of binaries in these simulations 
(since the softer binary is assumed to be disrupted in each interaction). In the absence of
evaporation through a tidal boundary, mass loss from the cluster comes almost entirely from
stars and binaries ejected through recoil following an interaction. The mass loss rate therefore
increases with increasing binary fraction. At core collapse, the total mass loss fraction
 is about 5\%, 15\%, and
25\% for $\fb =0.02$, 0.1, and 0.2, respectively.  
However, while the total number of binaries in the
cluster continuously decreases, the remaining fraction of binaries at core collapse
appears to be roughly constant, around 0.2, independent of $\fb$. This is sufficient
to power many cycles of gravothermal oscillations after the initial core collapse, even 
for initial binary fractions as small as a few percent. For $\fb =0.2$ (Fig.~5), we 
were able to extend our numerical integration all the way to almost $\sim 10^3\,\trh$,
at which point several thousand binaries are still present in the central region of
the cluster (this timescale would of course vastly exceed a Hubble time for most Galactic
globular clusters!).

The bottom panels in Figures~3--5 show the evolution of several characteristic radii.
The most important is the core radius $\rc$ (recall that, by our definition, eq.~(\ref{eq:rc}), the central density
scales approximately as $\rho_0\propto \rc^{-2}$ since the central velocity dispersion is approximately
constant). Even in deep core collapse, the core radius of our models never decreases by more than a factor
$\sim 100$ (corresponding to an increase in the central density by $\sim 10^4$). Models with higher
binary fractions contract very little (see Fig.~5: the first and deepest core collapse corresponds to
a decrease in $\rc$ by less than a factor 10).
Also shown are the half-mass radii of the binaries $r_{\rm h,b}$ and of the single stars
$r_{\rm h,s}$.
The half-mass radius of the single stars always increases monotonically for these isolated clusters.
In contrast, the half-mass radius of the binaries tends to increase on average but shows a much more
complex behavior that depends strongly on the binary fraction and on the particular dynamical phase
in the evolution of the cluster.  The trend is for $r_{\rm h,b}$ to decrease during normal cluster
evolution, as the binaries mass segregate to the cluster core, and to increase dramatically
during core collapse, as the density of binaries in the core grows and the rate of 
binary--binary interactions grows more quickly than the rate of binary-single interactions.
This causes many softer
binaries in the core to be disrupted and many harder binaries to be ejected out of the core
through recoil.  For sufficiently low binary fractions (Figs.~3 and~4), $r_{\rm h,b}$
eventually becomes larger than $r_{\rm h,s}$ after core collapse. For high binary fractions (Fig.~5),
the binaries remain always much closer to the center of the cluster.

We now turn to a more detailed discussion of the Plummer model with 10\% binaries (Fig.~4), 
including a comparison with the \FP\ results of \gaoetal\ (see their Figs.~1--3), who
consider this their ``standard model.'' Qualitatively, our results are in very good 
agreement up to core collapse. After an initial phase
of contraction lasting $\sim 10\,t_{\rm rh}$, the core radius becomes nearly constant and
the cluster enters a long phase of quasi-thermal-equilibrium. This is the stable ``binary
burning'' phase, analogous to the main sequence for a star. During this phase, the rate of energy
production through interactions in the core is balanced by the rate at which energy flows out
in the outer halo, which continuously expands (in the absence of a tidal boundary). Core collapse
occurs rather suddenly at the end of this phase. \gaoetal\ find $\tcc/\trh \simeq 50$ for this model,
while we find $\tcc/\trh \simeq 70$.
This initial core collapse is followed by 
gravothermal oscillations, which are clearly still powered by primordial binary burning.
We were able to follow these oscillations accurately until $t/\trh \go 200$, while \gaoetal\ terminate their
calculation at $t/\trh \simeq 90$. 

Upon closer examination and quantitative comparisons, 
some more significant differences become apparent. First, we see that the initial 
contraction phase appears much deeper in the model of \gaoetal, with $\rc$ decreasing by almost 
an order of magnitude, while in our model the core contracts by a factor of about 3.
During the stable binary burning phase, the cluster also appears somewhat more centrally concentrated in
the model of \gaoetal. Just before core collapse, they find $r_{\rm c}/r_{\rm h}({\rm s})\simeq 0.01$,
while our value is $\simeq 0.04$. On the other hand, the rate
of binary burning and destruction is nearly the same in the two models. Compare, for example, the
evolution of $M_{\rm b}/M_{\rm b}(0)$ in Figure~4 to the same quantity plotted in Fig.~2a of \gaoetal.
Although there are slight differences in the shapes of the two curves, the reduction to 0.8 occurs 
after about $10\,t_{\rm rh}(0)$ in both cases, and the reduction to 0.5 after 
about $28\,t_{\rm rh}(0)$. By $t/t_{\rm rh}(0)\simeq 70$ the number of binaries has been reduced
to 0.2 of its initial value in both models. This agreement is especially surprising since in our model
this is still (just) before core collapse, while in \gaoetal's model several cycles of gravothermal 
oscillations have already occurred. 

There are several reasons to expect differences between our results and those of \gaoetal's
\FP\ simulations, even though our treatments of individual 
binary--single and binary--binary interactions 
are essentially identical.  

First, \gaoetal's representation of binaries
is in terms of a {\em separable} continuous distribution function in $E$, the orbital
energy in the cluster, and $\epsilon$, the internal energy of the binary.  In fact, there is
a strong and complex correlation between a binary's binding energy and its position
in the cluster (or equivalently its energy in the cluster), with harder binaries
concentrated near the cluster core (see \citealt{1992ApJ...389..527H} and Sec.~\ref{sec:binev}).  
We suspect that \gaoetal's choice
of a separable distribution function has the effect of reducing the energy generation
rate, since then proportionately more soft binaries will be chosen for binary
interactions --- interactions that predominantly liberate a constant fraction
of the total binding energy available (see Sec.~\ref{sec:binint} and \citealt{1975MNRAS.173..729H}).

Second, 1-D \FP\ results are known to differ from 2-D results in general
(most notably in the prediction of the rate of tidal stripping; see Paper II).
Enforcing isotropy in the stellar velocity distribution is likely to affect
the dynamics of the core around the time of collapse, when this distribution may
be changing rapidly and the increased interaction rate may be causing anisotropy.
Indeed, \citet{baumgardt2003} have recently found with $N$-body simulations
that anisotropy near the cluster center
becomes significant during core collapse.

Third, the only explicit dependence on $N$ in the \FP\ approach is through the Coulomb
logarithm, so, even though \gaoetal\ set $N=3\times10^5$ for their treatment of interactions, 
it is not clear in what sense their
results, which assume a smooth, continuous distribution function, correspond
to this particular value of $N$.

Finally, we point out that our results for the initial contraction and core size during 
the binary-burning
phase are in much better agreement with those of direct $N$-body simulations including primordial
binaries. Indeed, both \citet{1992MNRAS.257..513H} (see their Figs.~5 and~18) and 
\citet{1990ApJ...362..522M} (see their Fig.~1) 
find, as we do, that the cluster core contracts typically by
a factor of about three from its initial size.  \gaoetal, on the other hand, find 
core contraction by an order of magnitude, a direct indication that their method
underestimates the energy generation rate.

Perhaps a more significant difference between our results and those of 
\gaoetal\ is in the post-collapse evolution.  \gaoetal\ find much more frequent, erratic, and
deeper gravothermal oscillation cycles.  
Our model shows almost quasi-periodic oscillations with period $\sim 40\,t_{\rm rh}$
and peak-to-peak amplitude $r_{\rm c,max}/r_{\rm c,min}\sim 100$. 
Instead, \gaoetal\ find 7 oscillations of widely varying periods
between $t/t_{\rm rh}(0)=50$ and~90, and $r_{\rm c,max}/r_{\rm c,min}\sim 10^3$.
We believe this may possibly be due to differences in the numerical method of calculating
$N_c$ (for use in eq.~(\ref{eq:trlx})---see discussion in Sec.~\ref{sec:units}), 
although it is not clear from their paper what method they actually use.
To check the validity of our results near core collapse, we have examined more carefully 
the dynamics governing core re-expansion after collapse.
In Figure~7, we show the evolution of the temperature profile in the cluster
as the system undergoes a core collapse and rebound. 
The temperature in the cluster normally decreases outward everywhere, as in a
star in thermal equilibrium.
However, during deep core collapse, a ``temperature inversion'' develops  
for a short time. This temperature inversion is responsible for driving
the rapid re-expansion of the core, as energy is now flowing inward. 
This mechanism has been predicted theoretically for a long time
\citep{1983MNRAS.204P..19S, 1989MNRAS.237..757H},
and has been observed directly in recent $N$-body simulations
\citep{1996ApJ...471..796M}. However, to our knowledge, ours is the first numerical
demonstration of this effect for a cluster containing primordial binaries.
In all previous studies, the binaries were assumed to form dynamically
via three-body interactions or two-body tidal captures during deep core 
collapse. As noted in the introduction, these mechanism are now considered
unrealistic, as they most likely lead to stellar mergers (which were not taken into 
account in the previous studies).

\subsection{Tidally Truncated Clusters}\label{sec:truncclus}

We now present our results for more realistic, tidally truncated clusters.
Figures 8--11 show the evolution of King models with $W_0=7$ and initial binary
fractions $\fb=0.02$, 0.05, 0.1, and 0.2, respectively. Several striking differences 
with isolated clusters are immediately apparent. First, we see that the initial core contraction 
phase is absent. For $\fb \le 0.1$, the core radius decreases slowly and monotonically 
all the way to collapse. For higher binary fractions (Fig.~11), the core radius {\em increases\/}
slightly at first. This is simply because the initial binary burning rate in this model is close
to the rate needed to reach thermal equilibrium. The higher the initial binary fraction
and central density (see below), the stronger
the tendency for the core to expand initially instead of contracting. Second, 
core collapse\footnote{It should be noted that, in much of the literature on globular cluster
dynamics, the term ``core collapse'' is used to refer to the initial core contraction phase (which we
see here can actually be expansion instead), and what we call the binary burning phase is then
called the ``post-collapse'' phase. Clearly this terminology no longer makes sense, and should be abandoned.
What we call ``core collapse'' in this paper refers to the brief episodes of deep core collapse
at the onset of and during gravothermal oscillations.} 
or complete disruption always occurs
in less than about $45\,\trh$. For $\fb \ge 0.1$, the disruption time $\tdis$ {\em decreases\/}
with increasing binary fraction, and complete disruption occurs before any deep core collapse.
This is in contrast to the models with lower binary fractions (Figs.~8 and~9), where core collapse
followed by gravothermal oscillations (similar to those observed for isolated clusters in the 
previous section) occur before disruption. The possibility for a cluster to suffer complete 
disruption before core collapse is a qualitatively new behavior introduced by primordial
binaries. Indeed, {\em all\/} King models without binaries (and without stellar evolution)
reach core collapse before disrupting (see Paper~II and \citealt{1996NewA....1..255Q}).

Figures~12 and~13 show the evolution of King models with 10\% binaries but different values
of $W_0$. For the very centrally concentrated cluster with $W_0=11$ (Fig.~12), significant core
expansion occurs in the first few relaxation times (with $\rc$ increasing by about an order of 
magnitude). This is a more extreme example of the behavior already noted in Fig.~11.
The final evolution of this cluster is also peculiar: this is one of few examples (see Table~1) 
we encountered where 
the binaries are completely exhausted before the cluster disrupts. At $t/\trh\simeq 30$, about 20\% 
of the initial cluster mass remains in single stars, and the cluster undergoes deep core collapse.
Since there are no binaries left, and our simulations include no other source of energy,
no re-expansion can occur and we must terminate the calculation.

For a model with $W_0=3$, which has a much more nearly-uniform density profile 
initially, complete disruption occurs before core collapse even with much lower
binary fractions (see Table~1). For the model with $\fb=0.1$ shown in Fig.~13,
 disruption occurs at $t/\trh\simeq 15$. For comparison
a single-component $W_0=3$ King model undergoes core collapse at $t/\trh\simeq 12$ (Paper~II).
Also note how the core contracts throughout the evolution at a nearly-constant rate much
faster than in models with higher values of $W_0$ (compare, e.g., Fig.~10). Just before
final disruption, the core radius has decreased by about an order of magnitude from its
initial value.

Only a few small $N$-body simulations of tidally truncated clusters with primordial 
binaries have been reported previously 
\citep{1992MNRAS.257..513H, 1994ApJ...427..793M}.
Detailed comparisons are not possible because these studies assumed rather different 
initial models and the $N$-body results (for $N\sim1000-2000$) are very noisy.
However, we do see good qualitative agreement, with mass loss rates $\sim10$ times
larger than for isolated clusters, and complete disruption also 
observed after a few tens of initial half-mass relaxation times in all $N$-body 
simulations. 
We are not aware of any previous \FP\ simulations of tidally truncated clusters
with primordial binaries (\gaoetal\ considered only isolated Plummer models).

\subsection{Evolution of the Binary Population}\label{sec:binev}

In addition to affecting the global cluster evolution, binary interactions
also strongly affect the properties of the binaries themselves.  The study
of the evolution of a primordial binary population dates back to the seminal
work of \citet{1975MNRAS.173..729H}, but it is only recently that detailed numerical simulations
of large binary populations in globular clusters have been performed
\citep{1992ApJ...389..527H, 2000MNRAS.317..581G}.
We can use the results of our
Monte Carlo simulations to study the dynamical evolution of the binary population.

Fig.~\ref{fig:fb_evol} shows the evolution of the binary fraction, $\fb$,
in different regions of our evolving King models.  There is a clear trend for $\fb$ to increase
in the core and decrease in the halo with time, as well as a trend for
$\fb$ to grow more with smaller $W_0$. In spite of mass segregation and the tendency
for binaries to dominate the central region of a cluster (following the development of
the Spitzer instability; see Sec.~\ref{sec:isoclus}), core binary fractions rarely exceed
0.5 in our models. Thus the range of initial binary fractions we consider are at least
in rough agreement with the measurements of core binary fractions in globular clusters
today ($\sim0.1-0.4$; see Sec.~1). Note, however, that the present-day binary
fraction in the core of a cluster cannot be related simply to the cluster's
initial binary fraction, as it may depend in a complicated way on several initial parameters.
For example, we see that an initial $W_0=11$ model
with $\fb=0.1$ has, during most of its evolution, about the same core binary fraction
($\simeq 0.15$) as an initial $W_0=7$ model with $\fb=0.05$. In addition, recall that
our definition of $\fb$ in these simulations includes only the {\em hard binaries\/}. For reasonable
distributions of primordial binary separations, including several more decades on the soft
side, the true initial binary fraction in the cluster might have been $\sim2-3$ times larger
than our quoted value of $\fb$ (this is the reason why we did not consider values of
$\fb\go 0.3$, which could not be realistic, unless dynamics already plays an important
role during the process of star formation; see \citealt{2000prpl.conf..151C}).

Fig.~\ref{fig:gray_scale} shows the evolution of the primordial binary
population in a $W_0=7$ King model with $\fb=0.2$.  Each 2-D histogram shows
the distribution of binding energies (initially flat
in $\log\epsilon$) and radial positions in the cluster.
In addition to the clear tendency for mass segregation and hardening of the binaries,
we note the development of a strong correlation between hardness and
radial distribution: harder binaries tend to concentrate in the cluster core 
much more than softer binaries, in spite of having all the same mass (and in contrast to the 
fundamental assumption made in
the \FP\ calculations of \gaoetal). The general trends observed here are in good qualitative
agreement with the results of previous studies using more idealized models
\citep[see, e.g., their Fig.~25]{1992ApJ...389..527H, 2000MNRAS.317..581G}.

Near the end of the evolution shown in Fig.~\ref{fig:gray_scale}
(but already $\sim10\,\trh$ before complete disruption, when the cluster
still retains about 40\% of its initial mass), a particularly
striking situation develops where all the surviving binaries in the cluster core
are extremely hard. Recall that our initial upper limit on the binding energy 
of a binary ($\sim10^2\,kT$, where ``$kT$'' is the average kinetic energy of stars in the core) roughly
corresponds to contact for two solar-like stars. Therefore, most of the binaries remaining
after $\sim30\,\trh$, with binding energies now in the range $\sim10^2-10^3\,kT$, would have
merged if they contained solar-like stars (perhaps forming blue stragglers). Of course,
in a real cluster, many of these binaries could contain compact objects (most likely heavy white dwarfs  
and neutron stars) and would then have survived. 
We cannot address any of these issues here, since our simulations are clearly
too idealized, but we point out that globular cluster cores are indeed observed to contain
large populations of blue stragglers, WUMa binaries (eclipsing systems containing two 
main-sequence stars in a contact configuration; see, e.g., \citealt{2001ApJ...559.1060A}), 
and a variety of ``ultracompact'' binaries
containing neutron stars and white dwarfs (the most extreme example being perhaps
the ``11-minute'' X-ray binary in NGC 6624; see, e.g., \citealt{2000ApJ...530L..21D}).

\section{Summary and Comparison with Observations}

We have performed, for the first time, discrete simulations of globular
clusters with realistic numbers of stars and primordial binaries, using
our 2D Monte Carlo code with approximate analytical cross sections
for primordial binary interactions.

We have compared the use of cross sections with exact, dynamical integrations
of binary--single encounters, and find that the agreement between the two
methods is strong, although our current implementation of the cross sections,
based on the \FP\ study by \citet{1991ApJ...370..567G}, tends to overestimate slightly the energy
generation rate.  Consequently, models that use cross sections tend to
overestimate core collapse times for clusters in which binary--single
interactions dominate.  However, we find that binary--binary interactions dominate
the energy generation for $\fb\gtrsim0.03$, a result that is in quantitative
agreement with a simple Spitzer-type stability criterion applied to the component of 
binaries.

We have studied the evolution of isolated clusters with varying binary
fractions, and have found that the presence of even a small fraction of
binaries is sufficient to delay significantly the onset of core collapse.  
Isolated clusters
with a binary fraction greater than about 0.1--0.2 can avoid core
collapse for as much as $\sim 10^2-10^3\,\trh$.  We find a simple linear relation
between the core collapse time of a cluster, $\tcc$, and the core collapse
time of the same cluster with binary interactions turned off
(but mass segregation still present), denoted
$\tcc(\fb=0)$, given by $\tcc \simeq \tcc(\fb=0) \times (75\,\fb + 1)$.
We have compared our results with those of \citet{1991ApJ...370..567G}, and find
reasonable agreement, with nearly identical rates of binary
burning and destruction.  \citet{1991ApJ...370..567G}, however, find a shorter
core collapse time, a deeper initial core contraction, and 
significantly more erratic behavior during the gravothermal oscillation
phase.  We attribute the differences primarily to their neglect of the strong 
correlation between binary hardness and spatial distribution in the cluster, as well as 
fundamental differences between their 1-D \FP\ method and our 2-D Monte Carlo method.
Our results for the initial core contraction are in much better agreement with those
of direct $N$-body simulations. In addition, we have presented
the first numerical demonstration of the theoretically predicted
temperature inversion powering re-expansion after
core collapse and gravothermal oscillations for a cluster with primordial binaries. 

We have also considered more realistic, tidally truncated King models.
We have found that the initial core contraction phase is absent in these systems, 
or replaced by an initial expansion of the core, 
and that core collapse or complete tidal disruption always occurs
in less than $\sim 50\,\trh$.  For a binary fraction $\gtrsim 0.1$, the disruption
time {\em decreases} with increasing binary fraction, and complete disruption
occurs before any deep core collapse.  The possibility for a cluster to suffer
complete disruption before core collapse is a qualitatively new behavior introduced
by primordial binaries.  Our results are in good qualitative agreement with previous
studies of tidally truncated clusters containing primordial binaries.

We have already argued in Section~\ref{sec:binev} that our results are in general
agreement with current determinations of binary fractions in globular cluster cores,
typically $\sim0.1-0.4$. We now briefly consider our basic predictions for the
structural parameters of clusters during the binary burning phase and compare them
to the observed structural parameters of globular clusters.
While our models are clearly far too idealized for any detailed comparison, it
is useful to examine at least the most fundamental structural parameters: the core
radius $r_{\rm c}$, the half-mass radius $r_{\rm h}$, and, for tidally truncated
clusters, the tidal radius $r_{\rm t}$. Since the overall scale is largely irrelevant
(although it could in principle be set by relating the maximum binding energy of the binaries to
a stellar radius), we consider only the two ratios $r_{\rm h}/r_{\rm c}$ and
$r_{\rm t}/r_{\rm c}$ (or, equivalently, the {\em concentration
parameter\/} $c\equiv \log (r_{\rm t}/r_{\rm c})$ often derived by observers
using King model fits to photometric data).

Fig.~\ref{fig:conc_evol} shows
the evolution of $r_{\rm h}/\rc$ and the concentration parameter $c$
for several King models (see Sec.~\ref{sec:truncclus}).  Fig.~\ref{fig:obs_conc}
shows distributions of $r_{\rm h}/r_c$ and $c$ for Galactic globular
clusters, with data taken from the compilation of \citet{1996AJ....112.1487H}.  The top
panel shows a histogram of $r_{\rm h}/\rc$ values for all Galactic globular clusters, including
those classified observationally as ``core-collapsed\footnote{It is unclear what the relation
is, if any, between this observational classification --- based on the absence of a good
fitting King model --- and the various theoretical definitions of core collapse used in
the literature on dynamical modeling.}.''
The middle and bottom panels show the distributions of observed values for
$r_{\rm h}/\rc$ and $c$, respectively,
with the ``core-collapsed'' clusters excluded.  First, in
Fig.~\ref{fig:conc_evol}, note the tendency for the concentration parameter in clusters
with reasonable binary fractions ($\fb\gtrsim 0.05$) to
remain around, or even converge to, $c\simeq 1.5$.  This is in reasonable agreement with
the bottom panel of Fig.~\ref{fig:obs_conc}, which shows the
observed distribution also centered around 
$c\simeq 1.5$.  The top panel in Fig.~\ref{fig:conc_evol} shows
most clusters in the binary burning phase with $r_{\rm h}/\rc \lesssim 10$, also in quite 
reasonable agreement with the observed distribution
(Fig.~\ref{fig:obs_conc}~b), although the observed peak around $r_{\rm h}/\rc\simeq 2$
would require that most initial models be less centrally concentrated than our $W_0=7$ King models. 
We also note that both $c$ and $r_{\rm h}/\rc$ increase significantly,
and sometimes dramatically, for clusters approaching tidal disruption.
Thus the suggestion from our results might be that clusters classified observationally
as ``core-collapsed'' are those in the last few relaxation times before destruction in the Galactic
tidal field. Most of our King models appear to spend roughly the last 10--20\% of their lives
with $r_{\rm h}/\rc \go 10$, or $c \go 2$, again not too different from the observed 
fraction of ``core-collapsed'' clusters in our Galaxy (see Fig.~\ref{fig:obs_conc}~a).
Of course a more serious comparison should take into account real cluster ages and the
distribution of initial values for $\trh$, which is rather uncertain.

We are currently in the process of implementing exact dynamical integrations
to handle binary--binary interactions more accurately in our simulations, using
{\tt Fewbody}, a new small-N integrator we have written.  This integrator performs
automatic classification of outcomes, automatic stability analysis of arbitrarily
large hierarchies, autonomous integration termination, and stellar collisions.

\acknowledgements
We are very grateful to Douglas Heggie, Steve McMillan, Simon Portegies Zwart,
and Saul Rappaport 
for many helpful discussions. The three-body integrator used
in this work is part of the {\tt Starlab} software package developed by Piet Hut,
Steve McMillan and Simon Portegies Zwart.
This work was supported by NASA ATP Grant NAG5-12044 and NSF Grant AST-0206276. Some of our
numerical simulations were performed on parallel supercomputers at Boston University and 
NCSA under National Computational Science Alliance Grant AST980014N.  


\bibliographystyle{apj}
\bibliography{apj-jour,main}

\begin{deluxetable}{lrrrr}
\tabletypesize{\scriptsize}
\tablecaption{Model parameters and results. \label{tab:1}}
\tablewidth{0pt}
\tablehead{
\colhead{Model} & \colhead{$f_b$} & \colhead{$t_{\rm cc}\,[{\rm or}~t_{\rm dis}]/\trh$} &
\colhead{$M(t_{\rm cc})/M(0)$} & \colhead{$M_b(t_{\rm cc})/M_b(0)$}
}
\startdata
Plummer        &  2\% & 22   & 0.94  & 0.20  \\
Plummer        &  5\% & 39   & 0.90  & 0.14  \\
Plummer        & 10\% & 72   & 0.83  & 0.17  \\
Plummer        & 20\% & 120  & 0.75  & 0.28  \\
Plummer        & 30\% & 180  & 0.70  & 0.35  \\
King, $W_0= 3$ &  2\% & 15   & 0.25  & 0.04  \\
King, $W_0= 3$ &  5\% & 17   & 0.00  & 0.00  \\
King, $W_0= 3$ & 10\% & 15   & 0.00  & 0.00  \\
King, $W_0= 3$ & 20\% & 14   & 0.00  & 0.00  \\
King, $W_0= 5$ &  2\% & 18   & 0.42  & 0.07  \\
King, $W_0= 5$ &  5\% & 22   & 0.22  & 0.02  \\
King, $W_0= 5$ & 10\% & 25   & 0.00  & 0.00  \\
King, $W_0= 5$ & 20\% & 23   & 0.00  & 0.00  \\
King, $W_0= 7$ &  2\% & 17   & 0.67  & 0.14  \\
King, $W_0= 7$ &  5\% & 30   & 0.41  & 0.04  \\
King, $W_0= 7$ & 10\% & 50   & 0.00  & 0.00  \\
King, $W_0= 7$ & 20\% & 42   & 0.00  & 0.00  \\
King, $W_0=11$ &  2\% & 10   & 0.71  & 0.19  \\
King, $W_0=11$ &  5\% & 20   & 0.43  & 0.05  \\
King, $W_0=11$ & 10\% & 30   & 0.20  & 0.01  \\
King, $W_0=11$ & 20\% & 38   & 0.00  & 0.00  \\
\enddata
\end{deluxetable}

\clearpage 

\clearpage 
\begin{figure}[t]
\plotone{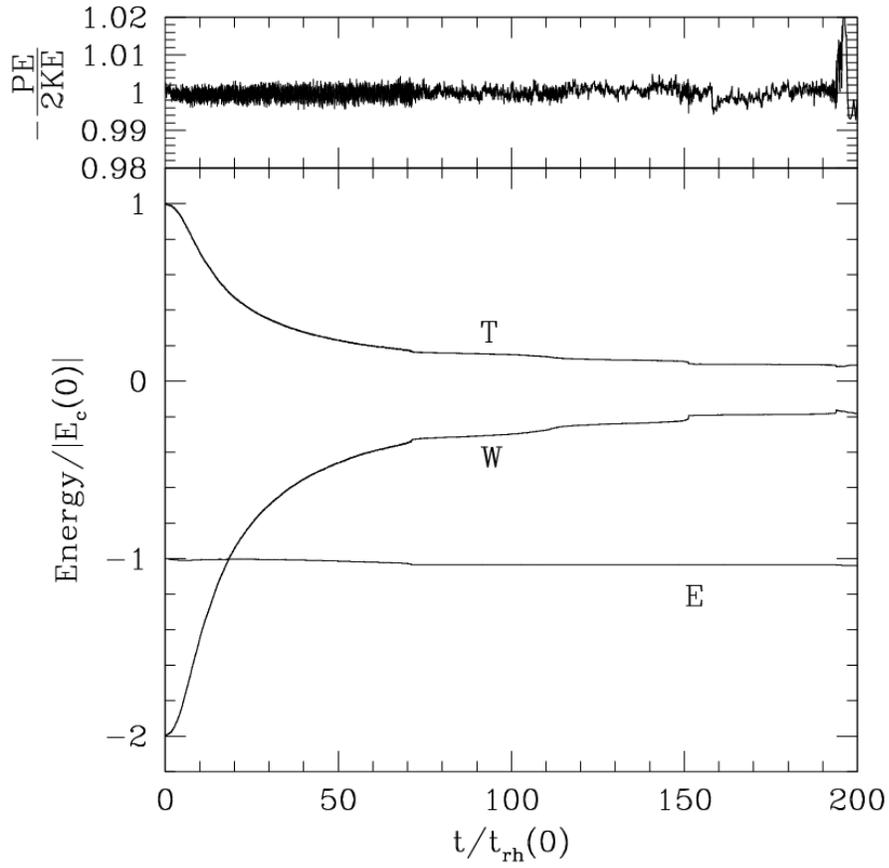}
\begin{center}
\caption{Evolution of the kinetic energy $T$, potential energy $W$ and total
conserved energy $E$, as well as the virial ratio $2T/|W|$, for a Plummer model with 
$N=3\times 10^5$ stars and 10 \% binaries initially. The cluster remains very close to virial 
equilibrium throughout the integration, and energy conservation is maintained to within
a few percent. 
Note that $E$ is corrected for both the energy lost through evaporation, and the energy
gained through binary--binary and binary--single interactions, so that $E\ne T+W$ (except at
$t=0$). The true total energy of the cluster, $T+W$, increases significantly over time as
a result of these interactions. Here we show the quantity that should be conserved, which we
monitor (in addition to the virial ratio) for quality control purposes in all our calculations. 
\label{fig:Evir}}
\end{center}
\end{figure}

\clearpage 
\begin{figure}[t]
\plotone{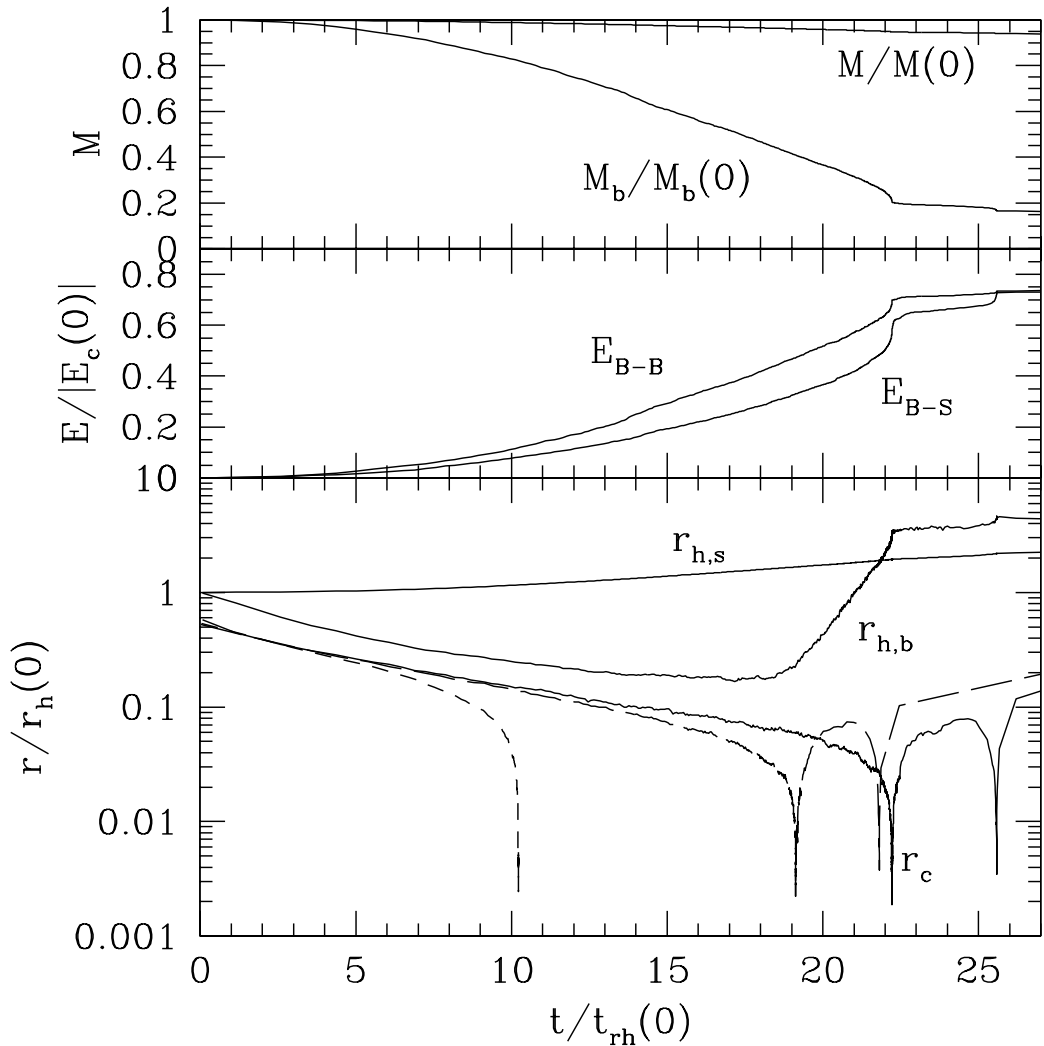}
\begin{center}
\caption{Evolution of an isolated Plummer model with $N=3\times 10^5$
stars and 2\% primordial binaries initially.  
The top panel shows the total cluster mass $M$ and the total mass $M_b$ in binaries. 
The middle panel shows the energy
released through binary--binary and binary--single interactions, in units of the initial
binding energy of the cluster. The lower panel shows the core radius $r_c$ of the cluster, 
the half-mass radius $r_{\rm h,s}$ of single stars, and the half-mass radius $r_{\rm h,b}$ 
of binaries (solid lines). For comparison,
the core radius of an equivalent Plummer model with 2\% primordial binaries
but with all interactions turned off is also shown (short-dashed line).
We see that even a primordial binary fraction as small as 2\% can significantly delay
core collapse, with $t_{\rm cc}$ increasing by more than a factor 2 in this case.
Also shown for comparison and testing is the core radius of an equivalent model where 
the binary--single interactions were computed with direct three-body integrations instead 
of cross sections (long-dashed line). Here again we note that the model based on direct 
integrations collapses slightly earlier.
\label{fig:plummer2}}
\end{center}
\end{figure}

\clearpage 
\begin{figure}[t]
\plotone{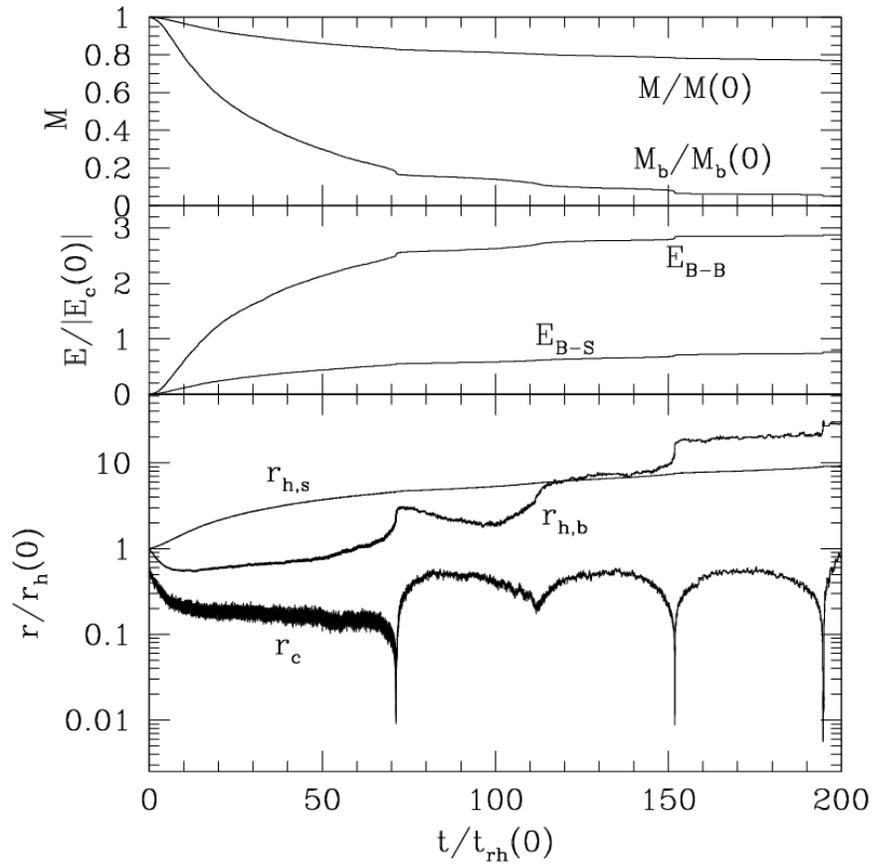}
\begin{center}
\caption{Same as Fig.~\ref{fig:plummer2}, but for a model with a 10\% primordial binary
fraction.  Here the energy generated from binary--binary interactions
clearly dominates that from binary--single interactions.
We see that an isolated cluster with 10\% binaries can be supported 
against collapse for about $70\, t_{\rm rh}$. Several cycles of gravothermal oscillations
powered by primordial binaries are seen after the initial collapse. The oscillations in this
case appear quasi-periodic with a period of roughly $50\,\trh$.
\label{fig:plummer10}}
\end{center}
\end{figure}

\clearpage 
\begin{figure}[t]
\plotone{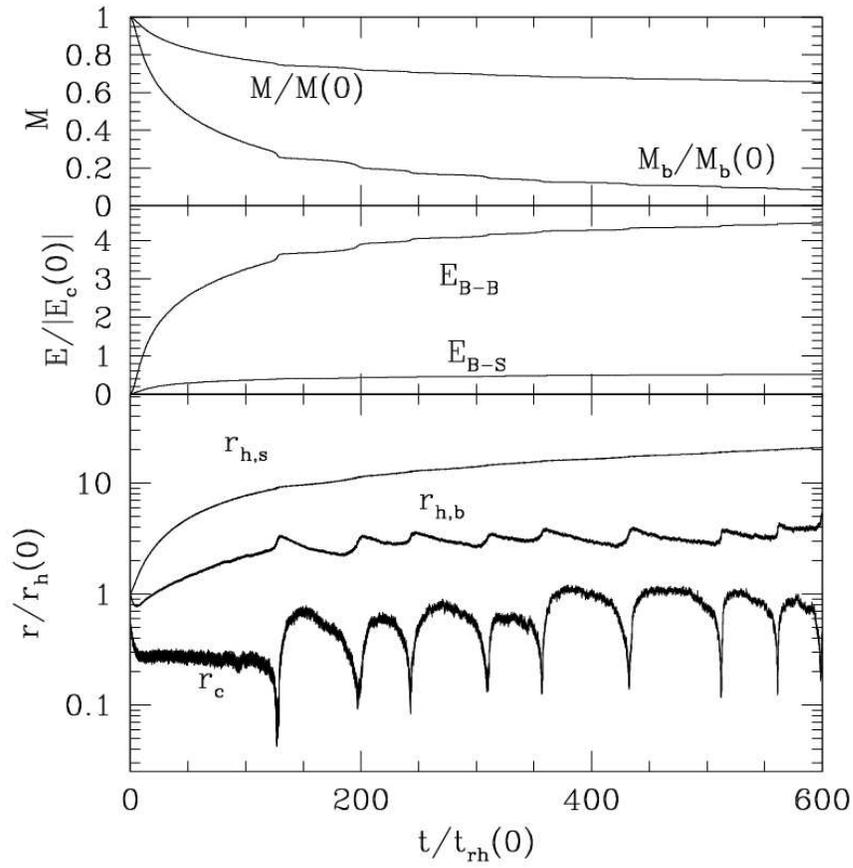}
\begin{center}
\caption{Same as Fig.~\ref{fig:plummer2}, but for a model with a 20\% primordial
binary fraction.  
Here the energy generated from binary--binary interactions is even more clearly dominant.
The cluster is initially supported 
against collapse for about $125\, t_{\rm rh}$. After the first core collapse,
gravothermal oscillations powered by primordial binaries continue up to 
$\sim 10^3\,t_{\rm rh}$. By that time the total number of binaries has been 
reduced by a factor $\sim10$, but the primordial binary reservoir is still not exhausted.
\label{fig:plummer20}}
\end{center}
\end{figure}

\clearpage
\begin{figure}[t]
\plotone{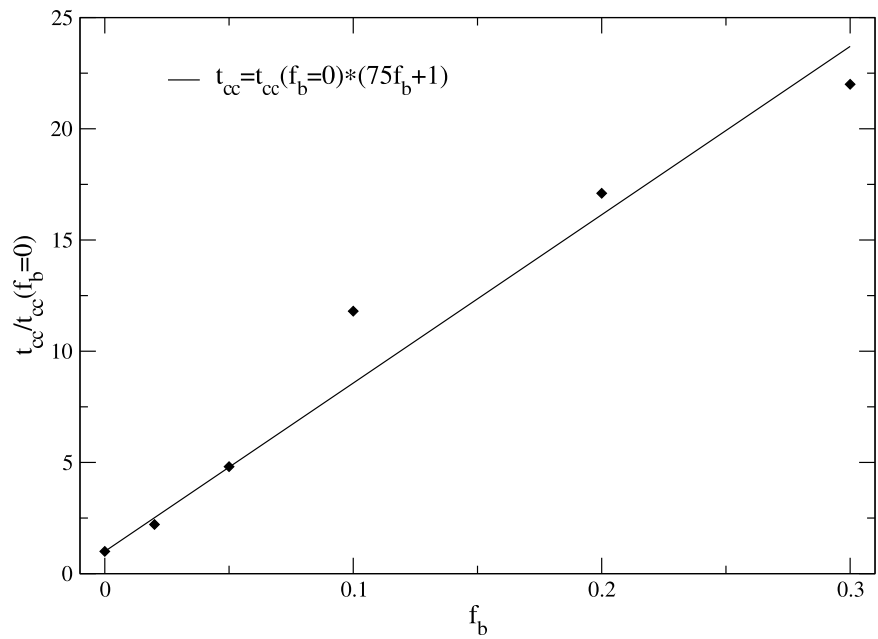}
\begin{center}
\caption{Ratio of core collapse times with and without binary interactions as a function of 
the initial binary 
fraction $f_{\rm b}$ for the Plummer models. The solid line shows a simple linear fit.
\label{fig:tcc_ratio}}
\end{center}
\end{figure}

\clearpage
\begin{figure}[t]
\plotone{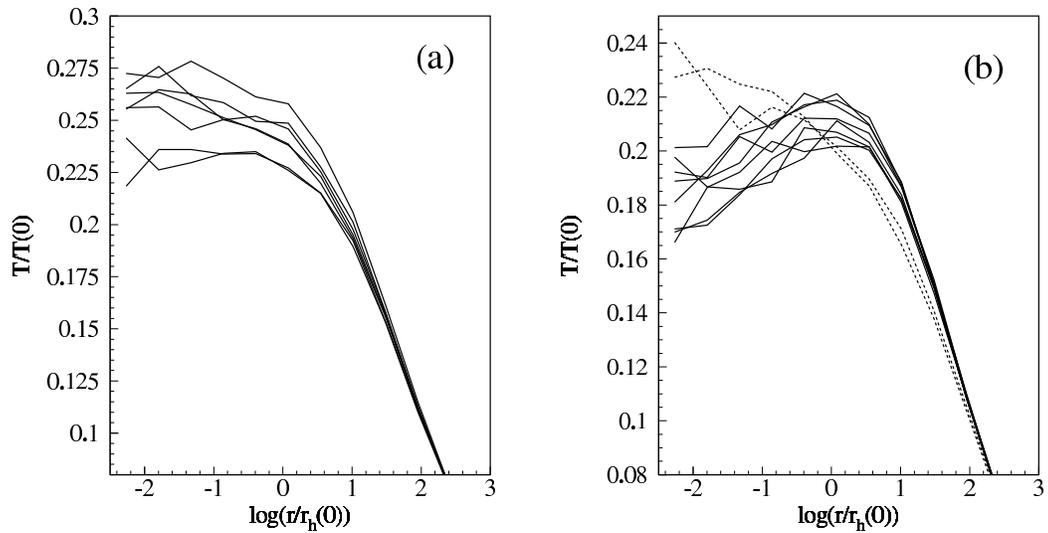}
\begin{center}
\caption{Evolution of the temperature profile near core collapse at $t\simeq 125\, t_{\rm rh}$
for an isolated Plummer model with $N=3\times 10^5$ stars and 20\% primordial binaries initially
(same model shown in Fig,~\ref{fig:plummer20}). The temperature is defined by $3kT=m\sigma_{\rm c}^2$.
A temperature inversion is clearly associated with re-expansion after core collapse.
The profiles are truncated very near the center where the number of stars is small and the statistical noise becomes too large. The profiles during 
contraction (a), evolving from top to bottom, are separated by about $0.1\,t_{\rm rh}(0)$, 
and the profiles during re-expansion (b), from bottom to top,
are roughly $0.05 t_{rh}(0)$ apart. The profiles shown by dotted lines in (b) are
about $4\,t_{\rm rh}(0)$ and $5\,t_{\rm rh}(0)$ after core collapse, indicating that the temperature inversion quickly disappears after re-expansion.\label{tempinvfig}}
\end{center}
\end{figure}

\clearpage 
\begin{figure}[t]
\plotone{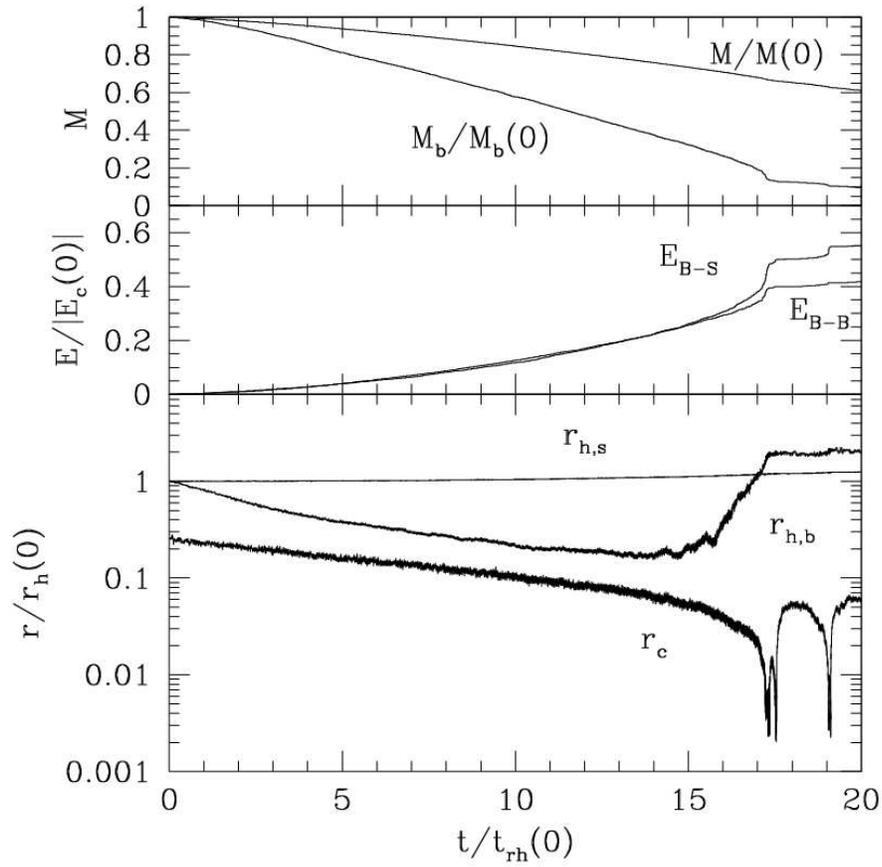}
\begin{center}
\caption{Evolution of a tidally truncated $W_0=7$ King model with $N = 3\times 10^5$
stars and 2\% primordial binaries. Conventions are as in Fig.~\ref{fig:plummer2}.
Compared to an isolated Plummer model with the same number of stars and binaries,
the evolution of this tidally truncated cluster to core collapse is only slightly faster.
\label{fig:king2}}
\end{center}
\end{figure}

\clearpage 
\begin{figure}[t]
\plotone{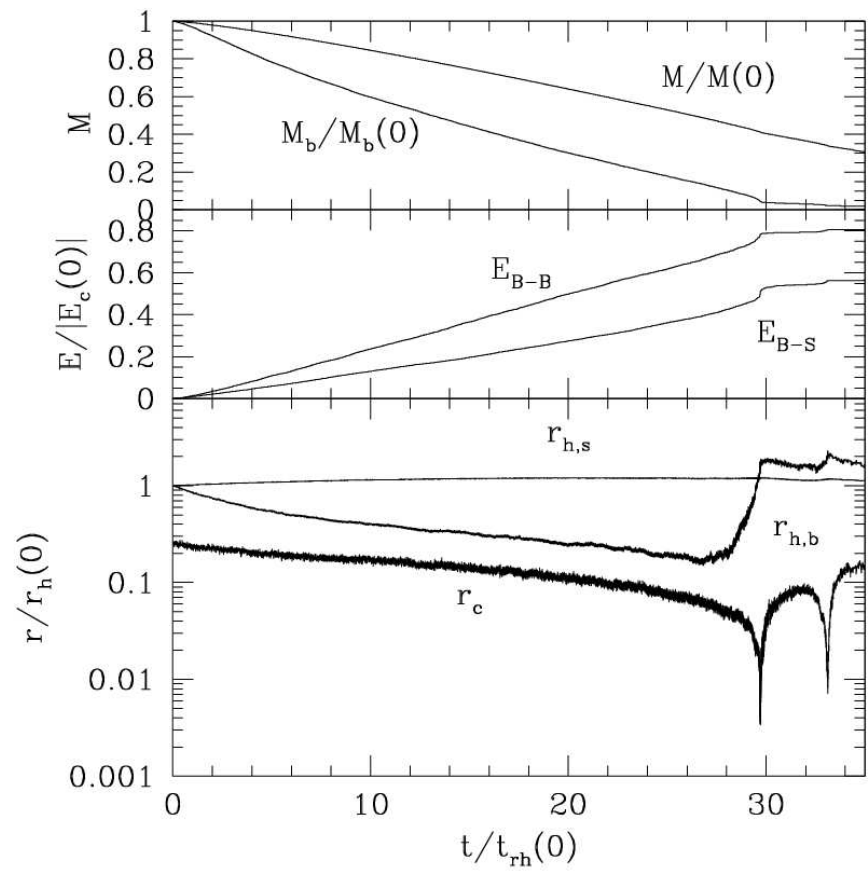}
\begin{center}
\caption{Same as Fig.~\ref{fig:king2}, but for a model with a 5\% primordial
binary fraction initially.
\label{fig:king5}}
\end{center}
\end{figure}

\clearpage 
\begin{figure}[t]
\plotone{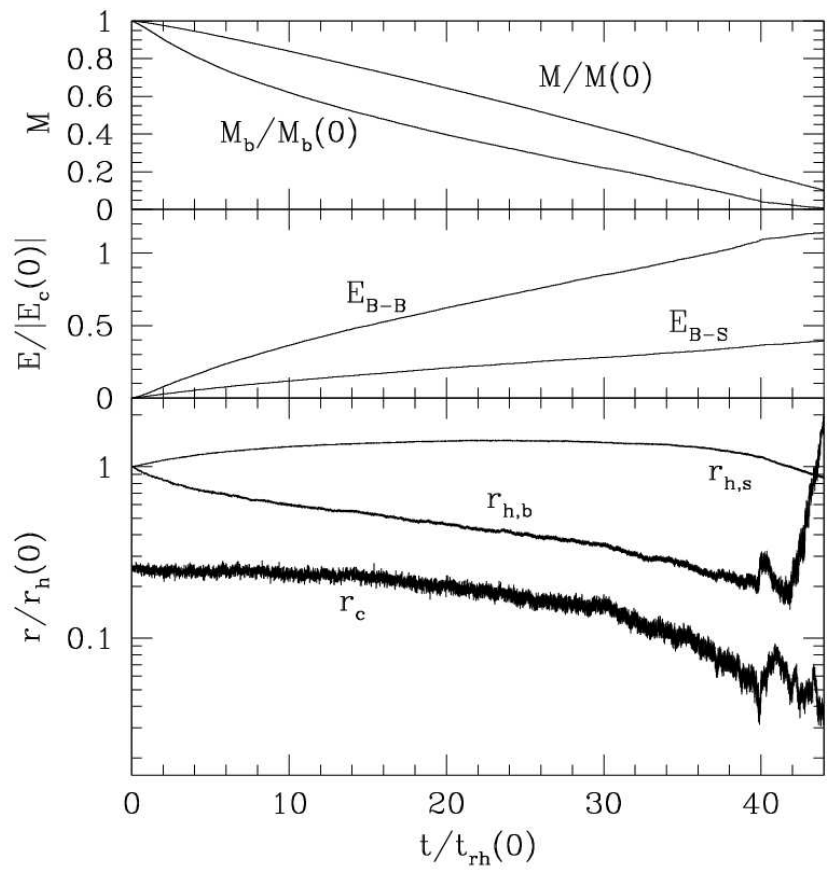}
\begin{center}
\caption{Same as Fig.~\ref{fig:king2}, but for a model with a 10\% primordial
binary fraction initially. Here complete disruption occurs before core collapse.
Note also the absence of any core contraction initially.
\label{fig:king10}}
\end{center}
\end{figure}

\clearpage 
\begin{figure}[t]
\plotone{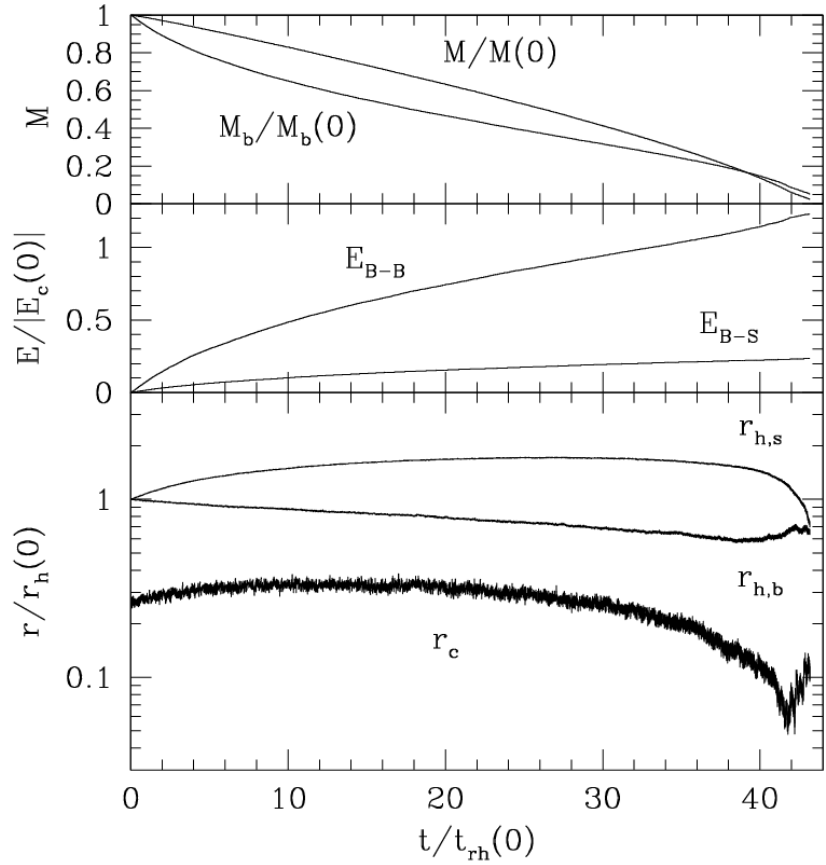}
\begin{center}
\caption{Same as Fig.~\ref{fig:king2}, but for a model with a 20\% primordial
binary fraction initially. Note the initial {\em expansion\/} of the core.
Here also complete disruption occurs before core 
collapse. The apparent re-expansion of the core radius after $t/\trh \simeq 42$
 is a numerical artifact caused by the very small number of stars left in the cluster
 (our method of calculating $\rc$ picks up stars well outtside the true core).
\label{fig:king20}}
\end{center}
\end{figure}

\clearpage 
\begin{figure}[t]
\plotone{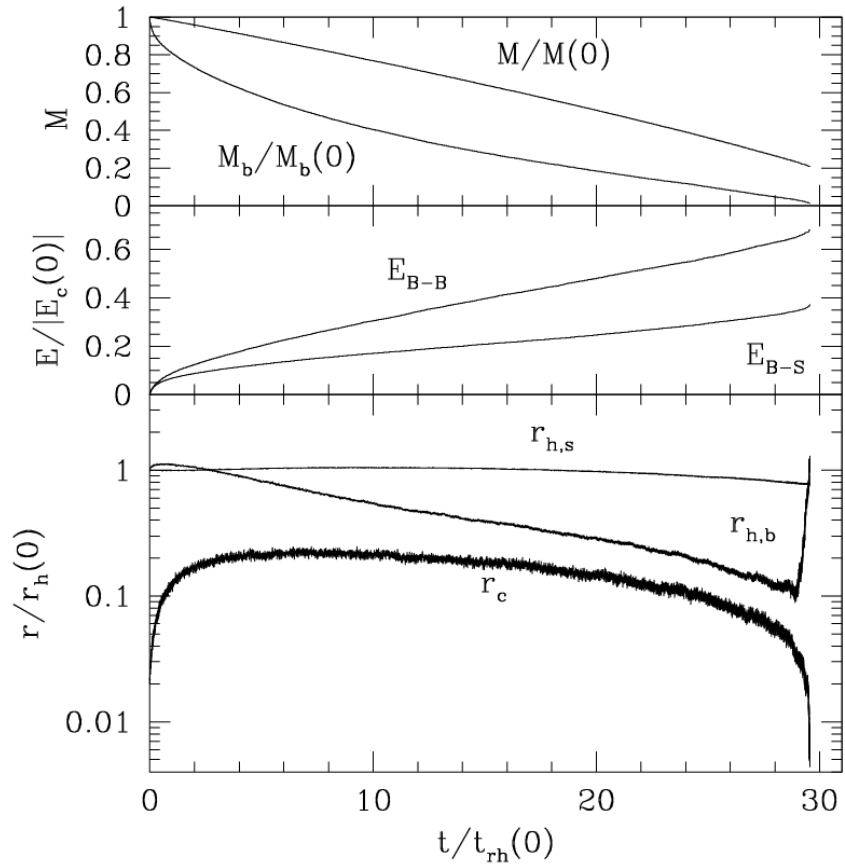}
\begin{center}
\caption{Same as Fig.~\ref{fig:king10}, but for a $W_0=11$ King model (with a 10\% primordial
binary fraction). This initially much more centrally concentrated model undergoes deep core collapse
as it runs out of binaries before complete disruption. Note also the significant initial
expansion of the core needed to reach quasi-equilibrium in a few $\trh$.
\label{fig:kingW11}}
\end{center}
\end{figure}

\clearpage 
\begin{figure}[t]
\plotone{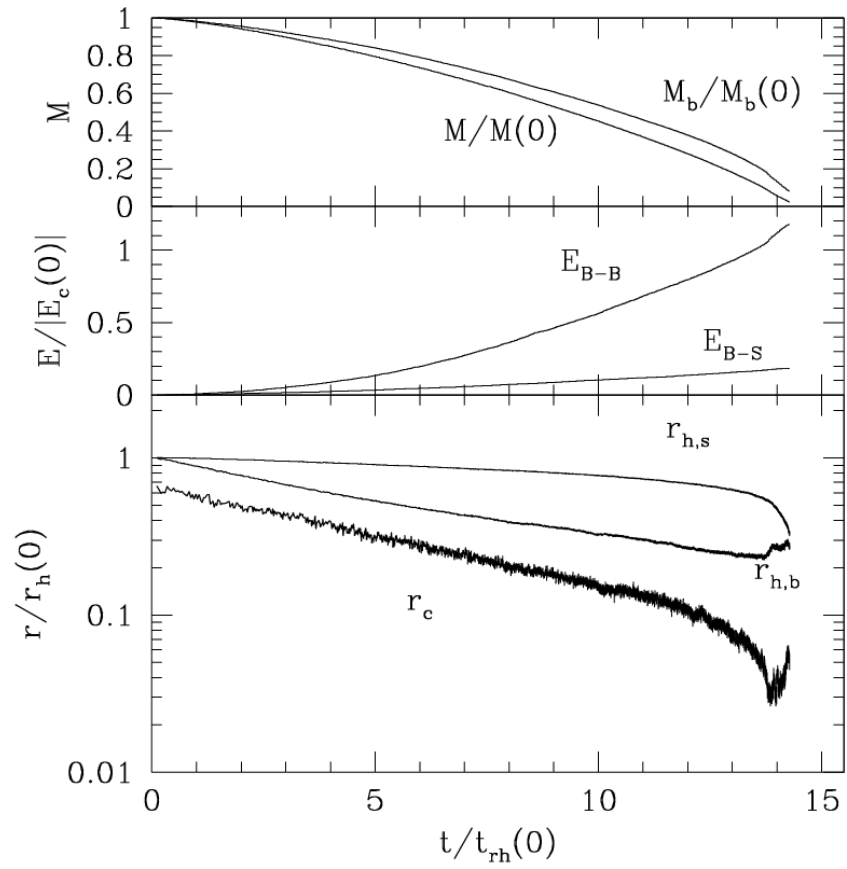}
\begin{center}
\caption{Same as Fig.~\ref{fig:king10}, but for a $W_0=3$ King model (with a 10\% primordial
binary fraction). This cluster is much less centrally concentrated initially and therefore,
as expected, it is disrupted before undergoing deep core collapse. Note, however, that significant
core contraction occurs throughout the evolution.
\label{fig:kingW3}}
\end{center}
\end{figure}

\clearpage 
\begin{figure}[t]
\plotone{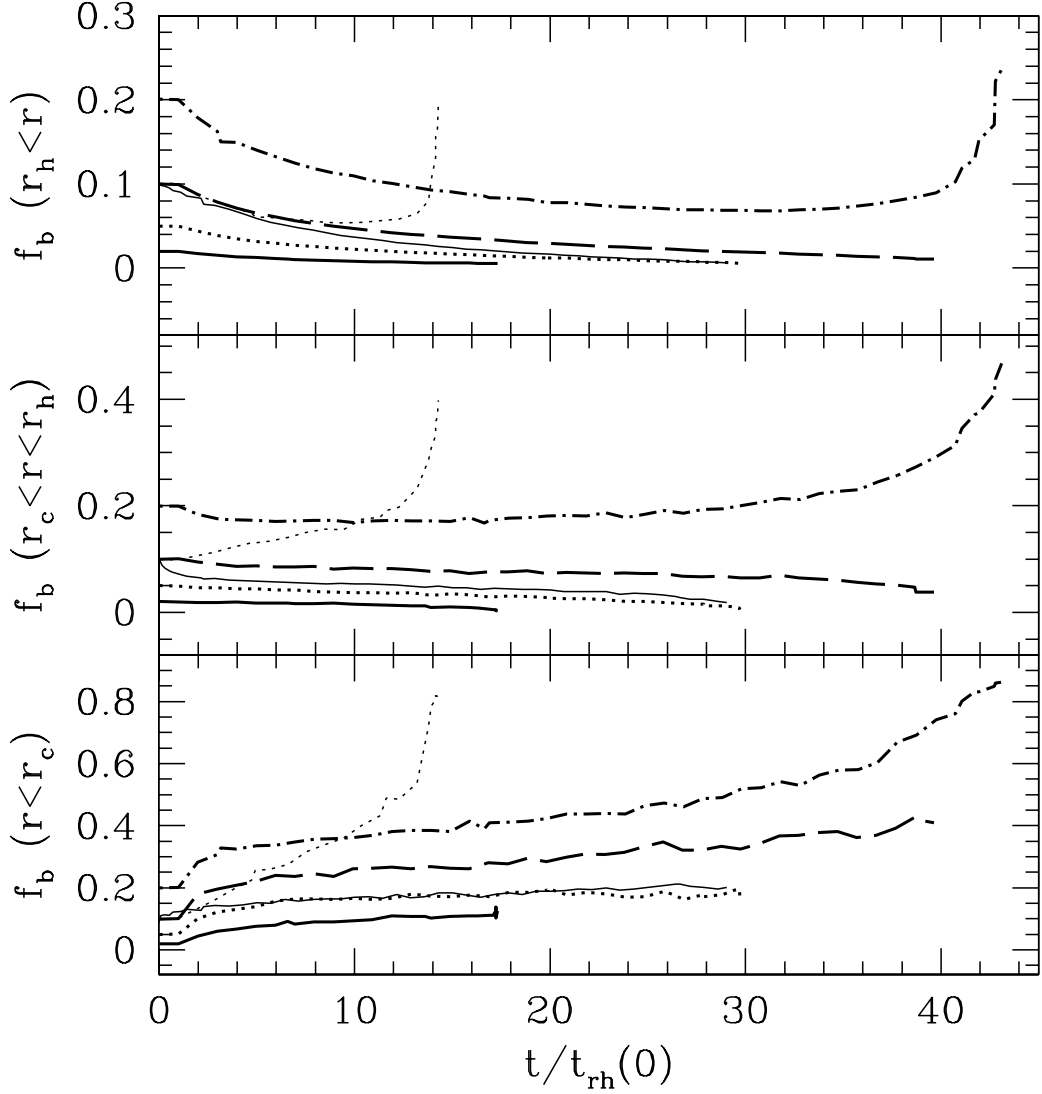}
\begin{center}
\caption{Evolution of the binary fraction $f_b$ in different regions for various King models.
The thin solid line is for a $W_0=11$ model with 10\% binaries
and the thin dotted line is for a $W_0=3$ model with 10\% binaries. The other lines are
for $W_0=7$ models with increasing binary fractions from bottom to top as in
Figs.~\ref{fig:king2}--\ref{fig:king20}. Binary fractions as high as 0.5--0.8
(but more typically $\simeq 0.1-0.2$) can be expected in cluster cores, while in the outer
halo one has typically $f_b \lesssim 0.1$.
\label{fig:fb_evol}}
\end{center}
\end{figure}

\clearpage 
\begin{figure}[t]
\plotone{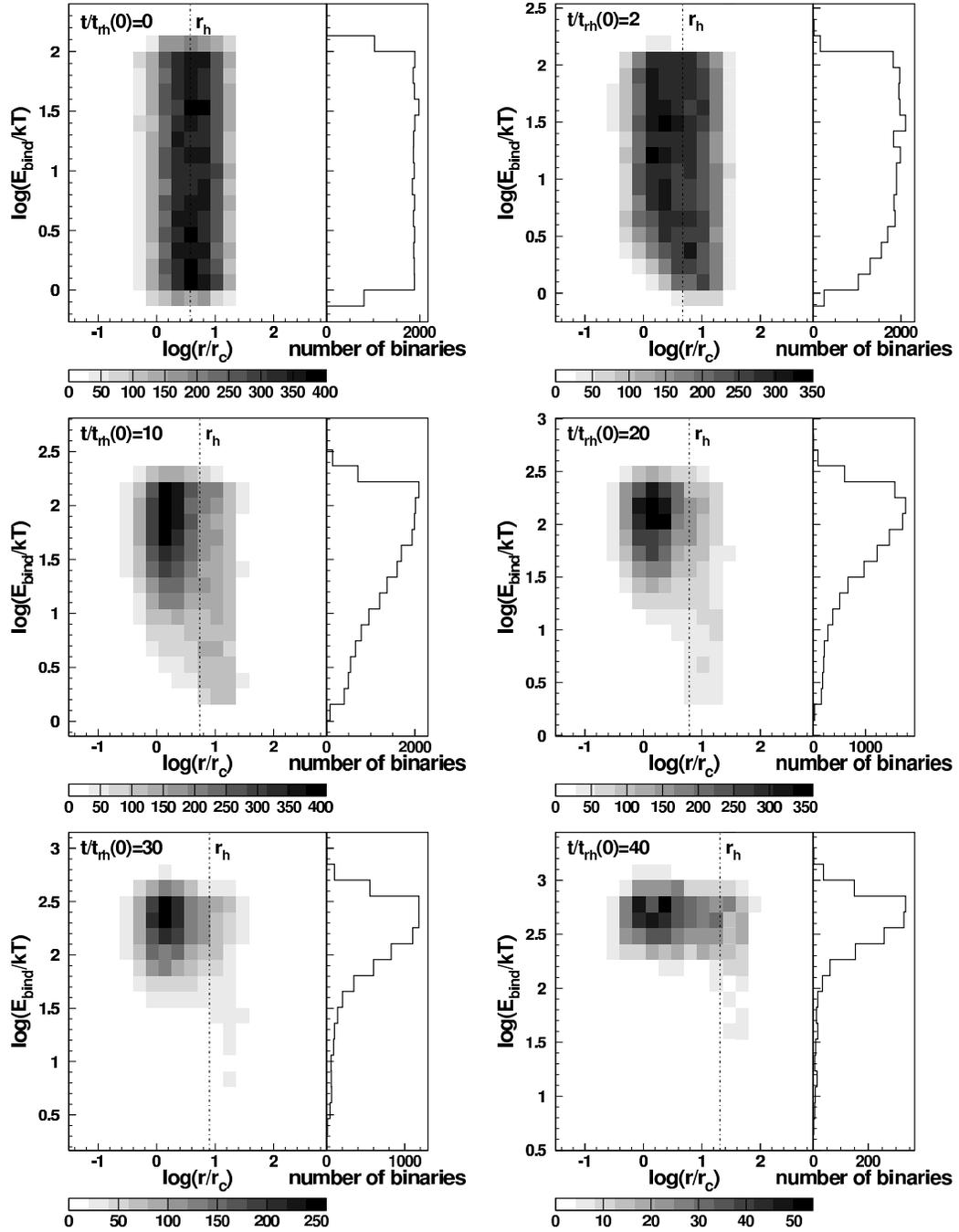}
\begin{center}
\caption{Evolution of the binary hardness and radial distributions for
a $W_0=7$ King model with 20\% binaries. Binaries undergo clear mass segregation,
and harden on average by about two orders of magnitude before cluster disruption.
\label{fig:gray_scale}}
\end{center}
\end{figure}

\clearpage 
\begin{figure}[t]
\plotone{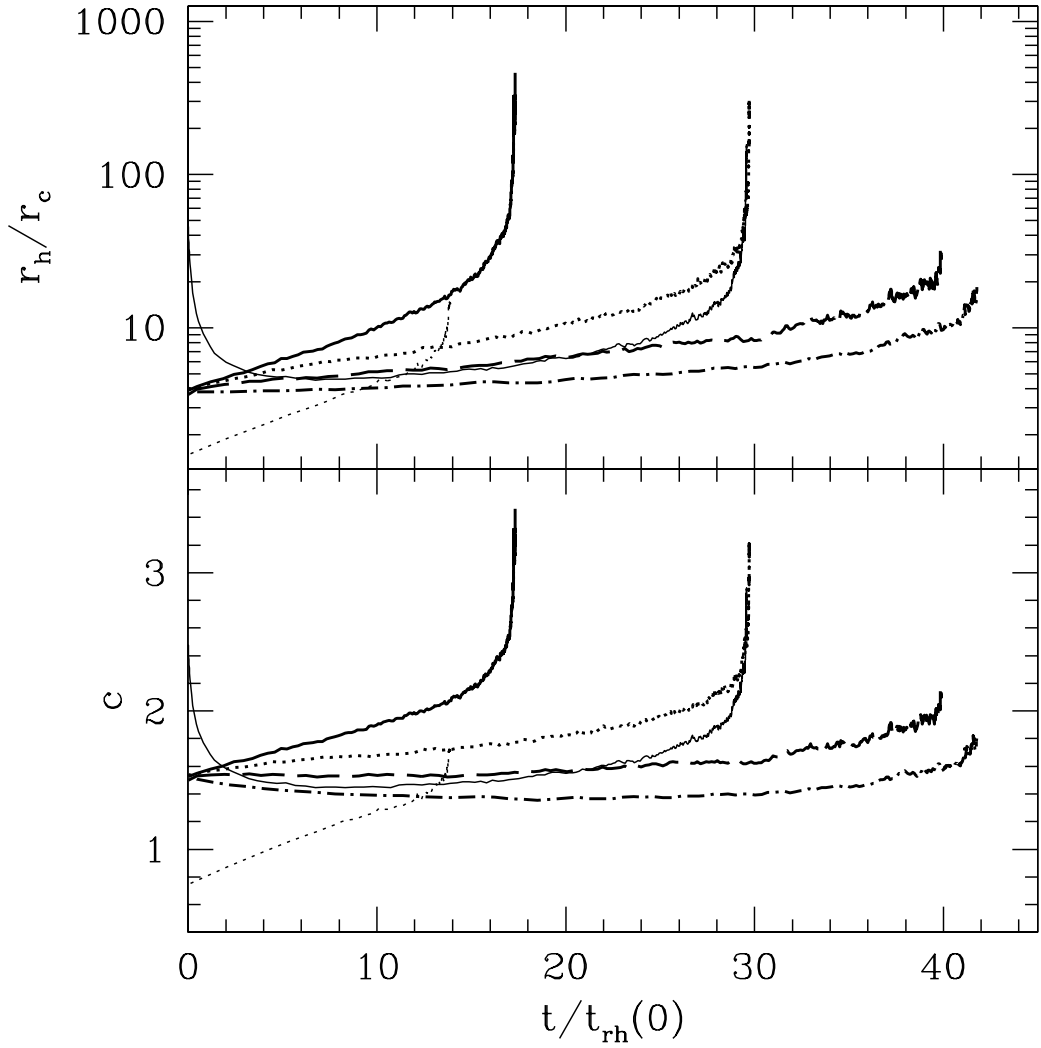}
\begin{center}
\caption{Ratio of half-mass to core radius and concentration parameter $c\equiv\log(r_t/r_c)$
for various King models.  The thin solid line is for a $W_0=11$ model with 10\% binaries
and the thin dotted line is for a $W_0=3$ model with 10\% binaries. The thick lines are
for $W_0=7$ models with increasing binary fractions from top to bottom, going from
2\%, 5\%, 10\%, to 20\%.
\label{fig:conc_evol}}
\end{center}
\end{figure}

\clearpage 
\begin{figure}[t]
\epsscale{0.67}
\plotone{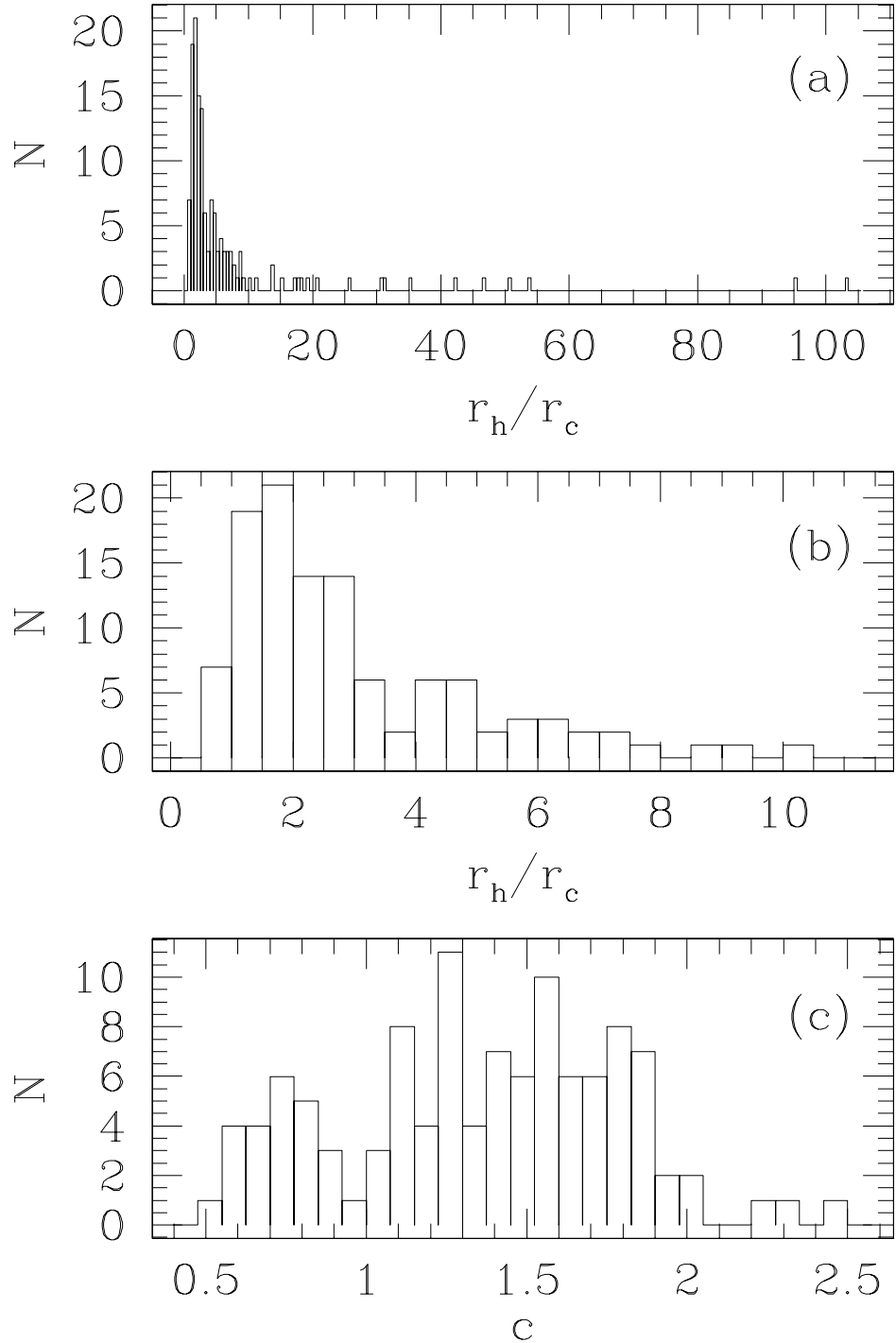}
\begin{center}
\caption{Observed distribution of $r_h/r_c$ and concentration parameter $c$ for Galactic globular clusters.
Clusters classified observationally as ``core-collapsed'' are included in (a) and (b), 
but not in (c). The observed values of these basic structural parameters for the
non-``core-collapsed'' clusters are clearly in
general agreement with values predicted by our simple models for clusters supported by
primordial binary burning. 
\label{fig:obs_conc}}
\end{center}
\end{figure}

\end{document}